\documentclass[11pt,a4paper,openright]{article}
\usepackage{graphicx}
\usepackage[english]{babel}
\usepackage[T1]{fontenc}
\usepackage[latin2]{inputenc}
\usepackage{amsmath}
\usepackage{amssymb}
\newcommand{\bea}{\begin{array}}
\newcommand{\eea}{\end{array}}
\newcommand{\be}{\begin{equation}}
\newcommand{\ee}{\end{equation}}
\newcommand{\ba}{\begin{eqnarray}}
\newcommand{\ea}{\end{eqnarray}}
\newcommand{\baw}{\begin{eqnarray*}}
\newcommand{\eaw}{\end{eqnarray*}}
\newcommand{\bt}{\begin{tabular}}
\newcommand{\et}{\end{tabular}}
\newcommand{\bc}{\begin{center}}
\newcommand{\ec}{\end{center}}
\newcommand{\ben}{\begin{enumerate}}
\newcommand{\een}{\end{enumerate}}
\newcommand{\bi}{\begin{itemize}}
\newcommand{\ei}{\end{itemize}}
\newcommand{\bmpage}{\begin{minipage}}
\newcommand{\empage}{\end{minipage}}
\newcommand{\bft}{\begin{footnotesize}}
\newcommand{\eft}{\end{footnotesize}}

\newcommand{\brak}{\langle}
\newcommand{\kket}{\rangle}

\def \dg{\dagger}
\def \df {{\sf d}}

\def \prt{\partial}
\def \inf {{\rm inf}\,}

\def \sup {{\rm sup}\,}

\def \Tr{{\rm Tr}\, }

\def\brak{\langle}
\def\kket{\rangle}

\def \al{\alpha}
\def \bet{\beta}
\def \del{\delta}
\def \Del{\Delta}

\def\la {\lambda}

\def\La {\Lambda}

\def\kgr{{\bf k} }

\def\0gr{{\bf 0} }

\def\sgr{{\bf s} }
\def\Sgr{{\bf S} }
\def\xgr{{\bf x} }
\def\ygr{{\bf y} }

\def\Bcal{ {\cal B}}

\def\Dcal{ {\cal D} }
\def\Ecal{ {\cal E} }

\def\Hcal{ {\cal H} }
\def\Ical{ {\cal I}}
\def\Jcal{ {\cal J}}

\def\Lcal{ {\cal L}}
\def\Mcal{ {\cal M}}

\def\Rcal{ {\cal R}}

\def\bsf{ {\sf b}}

\def\Cdb{ { \mathbb C}}

\def\Rdb{ { \mathbb R}}

\def\Zdb{ { \mathbb Z}}
\def\Tdb{ { \mathbb T}}

\newcommand{\limf}[2] {{\displaystyle\mathop{\lim}_{{#1}\to {#2}}  }}

\newcommand{\R}{\Rdb}

\begin{document}
\title{Operator reflection positivity inequalities and their applications
to interacting quantum rotors}
\author{\em Jacek Wojtkiewicz${}^\dg$,
Wies\l aw Pusz${}^\dg$,
 Piotr Stachura${}^*$
 \\
 ${}^\dg$ Department for Mathematical Methods in Physics,\\ Faculty of Physics, Warsaw University,\\
  Pasteura 5, 02-093 Warszawa, Poland
 \\
\\
 ${}^*$The Faculty of Applied Informatics and Mathematics,\\
Warsaw University of Life Sciences-SGGW,\\
ul. Nowoursynowska 159, 02-776 Warszawa, Poland
 }
 \maketitle
\abstract{In the Reflection Positivity theory and its application to statistical mechanical systems,
certain matrix inequalities play a central role. The Dyson-Lieb-Simon \cite{DLS}
 and Kennedy-Lieb-Shastry \cite{KLS2}
inequalities constitute prominent examples. In this paper we extend the KLS inequality
to the case where matrices are replaced by certain operators. As an application, we prove
the occurrence of the long range order in the ground state of two-dimensional quantum rotors.
}

\section{Introduction}
The {\em Reflection Positivity} notion has appeared in Quantum Field Theory in seventies
of the last century \cite{OS}.
Few years later, it has been applied to investigation of phase transitions
 in both classical \cite{FSS76}
and quantum \cite{DLS} lattice spin systems. The Reflection Positivity
turned out to be a very useful tool,
giving the first rigorous proofs of existence of phase transitions
in systems with continuous symmetry group.

The cornerstone of Reflection Positivity for quantum spin systems
 is the matrix inequality due to Dyson, Lieb and Simon (Lemma 4.1
in \cite{DLS}). Using this Lemma, authors proved
the existence of orderings in the XY as well as Heisenberg models in $d\geq 3$ and
for  sufficiently small temperature. Later on, this method has been extended to certain class of
{\em infinite dimensional} operators. This way, the existence of Long-Range Order has been proved
for $d\geq 3$ in the system of quantum interacting rotors  \cite{PaChor}.

Another direction of development of Reflection Positivity techniques was an examination of {\em ground
states} of quantum spin systems and orderings therein. It turned out that one can take certain
zero-temperature limit in the framework of
 the DLS method. This way, the appearance of Long Range Order has
been proved in XY and Heisenberg models in  $d=2$ \cite{NevesPerez}, \cite{Kubo}, \cite{KLS1}.
Later on, it turned out that such a proof can be done directly in the ground state,
with the use of another  matrix inequality, due to Kennedy, Lieb and Shastry (KLS)
 \cite{KLS2}. This inequality was further
generalized by Schupp \cite{Schupp}.

 It would be tempting to extend this inequality
to infinite-dimensional version, i.e. for certain class of  operators.
 However, to our best knowledge,
the operator version of the  KLS and Schupp(KLSS) inequalities, suitable for applications to ground states
of quantum interacting rotors has {\em not} been developed.

This opportunity inspired us to attempts to prove an operator analog of the KLSS inequalities.
It turned out to be possible, and this is one of two main results of our paper:
{\em extension of the KLSS matrix inequalities
 to certain class of infinite-dimensional operators}.
The second group of results which seems to be new are some applications.

The outline of the paper is as follows. In the Sec.~\ref{sec:KLSSop} we formulate the operator
version of the KLSS inequalities.
 The  application of this inequality is described in the Sec.~\ref{sec:LRO};
 it is the proof of the
ordering in ground state of $d\geq 2$ rotors (alternative proof to that presented in \cite{W1}).
The Sec.~\ref{sec:Summary} contains summary, conclusions and description of some open problems.
\section{KLS inequality and its extension for operators}
\label{sec:KLSSop}
\newtheorem{defi}{Definition}[section]
\newcommand{\dowl}{\hspace*{\fill}\rule{1ex}{1ex}\hspace*{1em}}
\newtheorem{tw}[defi]{Theorem}
\newtheorem{prop}[defi]{Proposition}
\newtheorem{lem}[defi]{Lemma}
\newtheorem{col}[defi]{Corollary}
\subsection{Kennedy, Lieb, Shastry and Schupp matrix inequalities. }
For convenience of the reader, and to show the idea of a proof without
operator-theoretic details, we present firstly the matrix version of KLSS inequality.

\begin{tw} {\rm \cite{KLS2}}
Let  $c, A, B$ be  $n\times n$ complex matrices,
$|c|:=\sqrt{c^*c} $ and $|c^*|:=\sqrt{c\,c^*}$ 
the moduli of $c$ and $c^*$ respectively.
 Then
\be
|\Tr\, c^*\,B\,c\,A^*|
\leq
 \frac{1}{2}\left[\,\Tr\,(|c|\,A\,|c|\, A^*) + \Tr\,(|c^*|\,B\,|c^*|\,B^*)\,\right].
\label{KLSorig}
\ee
\end{tw}
{\bf Sketch of the proof: }  At first let us note that by the polar decomposition theorem $c$ is of the form
$c=u\,|c|$,  where $u$ is a partial isometry. Since $u^*\, u \,|c|=|c|$ and $u\,|c|\,u^*$ is a positive  matrix, the polar decomposition of $c^*$ is of the form:
\be
c^* = |c|u^* = u^*\,u\,|c|\,u^* = u^*|c^*|.
\label{cstarM}
\ee
Taking adjoint we get
$\displaystyle 
c = u\,|c| = |c^*|\,u .
$
Therefore 
\be
u\sqrt{|c|} = \sqrt{|c^*|}\,u
\label{ucLcRuM}
\ee
according to functional calculus of positive hermitian  matrices and 
\be
c = \sqrt{|c^*|}\,u\,\sqrt{|c|}.
\label{cRucLM}
\ee
Now, let  $P$ and $Q$ be matrices introduced by formulae 
\be
P = u^*\,\sqrt{|c^*|}\,B^*\, \sqrt{|c^*|}\,u,\;\;\;\;\;
Q = \sqrt{|c|}\,A^* \,\sqrt{|c|}.
\label{PQM}
\ee
Let us remind that the trace functional defines the scalar product on the space of square matrices:
$(A|B):=\Tr\,A^*B.$  Therefore
\be
\left|\,{\Tr\,P^*Q}\,\right|\,\leq \,\sqrt{(\Tr\,P^*P)\,(\Tr\,Q^*Q )}\,\leq \,\frac{1}{2}[\,\Tr\,P^*P + \Tr\,Q^*Q \,]
\label{SchwPlusPQM}
\ee
due to the Schwarz inequality followed  by inequality between geometric mean  and arithmetic one. \\
Now using formula (\ref{ucLcRuM}) and  (\ref{cRucLM}) one can easily verify that
\[
P^*Q = u^*\,\sqrt{|c^*|}\,B\, \sqrt{|c^*|}\,u\,\sqrt{|c|}\,A^* \,\sqrt{|c|} =
\sqrt{|c|}\,u^*\,B\, c\,A^* \,\sqrt{|c|}
\]
and due to (\ref{cstarM})
\[
\Tr\,P^*\,Q = \Tr\, \left(\sqrt{|c|}\,u^*\,B\, c\,A^* \,\sqrt{|c|}\right) = \Tr\,(|c|\,u^*\,B\, c\,A^*)   = \Tr \,(c^*\,B\, c\,A^*).
\]
The module of it coincides with the left hand side of the
inequality (\ref{KLSorig}). \\
In the similar manner we compute the right hand side of (\ref{SchwPlusPQM}):
\[
P^*\,P = u^*\,\sqrt{|c^*|}\,B\, \sqrt{|c^*|}\,u\,u^*\,\sqrt{|c^*|}\,B^*\, \sqrt{|c^*|}\,u = u^*\,\sqrt{|c^*|}\,B\, |c^*|\,B^*\, \sqrt{|c^*|}\,u.
\]
Therefore
\be 
\Tr \,P^*\,P  = \Tr\,\left( u^*\,\sqrt{|c^*|}\,B\, |c^*|\,B^*\, \sqrt{|c^*|}\,u\right)=  \Tr \,(|c^*|\,B\, |c^*|\,B^*). 
\label{PstarPM}
\ee
By the similar reasoning  we get
\be
\Tr \,Q^*\,Q =  \Tr \,\left(\sqrt{|c|}\,A\, |c|\,\,A^*\,\sqrt{|c|}\,\right) = \Tr \,({|c|}\,A\,|c|\,A^*).
\label{QstarQM}
\ee
Combining (\ref{PstarPM}) and (\ref{QstarQM}) we obtain  the right hand side of inequality (\ref{KLSorig}).

\dowl
\begin{tw} {\rm \cite{Schupp}} The KLS inequality (\ref{KLSorig}) holds also for {\em rectangular} matrices,
i.e. $c$ is $n\times m$ matrix and matricies $A$ and $B$ are $m\times m$ and $n\times n$ respectively.
\end{tw}
{\bf Proof:} It is almost a repetition of the proof for KLS inequality and refers 
to the modified polar decomposition of the operator $c$ acting between $m$-dimensional and $n$-dimensional spaces.
In that situation  $|c|$, $|c^*|$ are positive matrices of dimensions $m\times m$ and
$n\times n$, respectively.
\dowl
\subsection{Operator version of the KLSS inequality}
The main goal of this subsection is to prove a generalized version of the KLSS  inequality:

\begin{tw}  Let $\Lcal$ and $\Rcal $ be separable Hilbert spaces, \,$A \in \Bcal(\Lcal)$, \ $B \in \Bcal(\Rcal)$  bounded operators acting on
$\Lcal$ and $\Rcal$ respectively.  Let $c: \Lcal \to \Rcal$
be a Hilbert-Schmidt operator  and   $|c|:=\sqrt{c^*c}$, and $|c^*|:=\sqrt{c\,c^*}$ be the corresponding moduli.
Then
\begin{enumerate}
\item  $|c|$ and $|c^*|$ are hermitean Hilbert-Schmidt operators acting on $\Lcal$ and $\Rcal$ respectively;
\item $c^*BcA^*$, \ $|c|A|c|A^*$ are trace-class operators on $\Lcal$ and $|c^*|B|c^*|B^*$ is
a trace-class operator on $\Rcal$;
\item the following inequality holds
\be
|\Tr\, c^*\,B\,c\,A^*|
\leq
 \frac{1}{2}\left[\,\Tr\,(|c|\,A\,|c|\, A^*) + \Tr\,(|c^*|\,B\,|c^*|\,B^*)\,\right]
\label{MainIneq}
\ee
\end{enumerate}
\end{tw}

\noindent
{\bf Remark}. In a finite-dimensional case, i.e. $\Lcal = \Cdb^N = \Rcal$ the above inequality
reduces to the matricial KLS inequality \cite{KLS2} and in more general finite dimensional situation
$\dim \Lcal \neq \dim \Rcal$ we obtain the result of Schupp \cite{Schupp}. \\

\noindent
{\bf Proof:} To prove our result we shall use some properties  of {\em Schatten ideals}  \cite{Schatten}, \cite{DunfordSchwartz2}, \cite{Kato} and 
now we  shall recall necessary results of the theory. 

Let $\Hcal$ be a separable Hilbert space, ${\mathcal C}\Bcal(\Hcal)$ - the set of compact operators on $\Hcal$.  
For a real number $p\geq 1$ the $p$-Schatten ideal is the set
\begin{equation*}
L^p(\Hcal):= \{ a \in {\mathcal C}\Bcal(\Hcal): \Tr (|a|^p) < \infty \}.
\end{equation*}
{\bf Remark.} Sometimes $L^p(\Hcal)$ is denoted as
$\Jcal_p$ \cite{RS1}, however, the actual notation corresponding to noncommutative $L^p$-spaces seems more natural to us.\\\\
For $a \in L^p(\Hcal)$ let us define:
\[
||a||_p := \left(\Tr\,|a|^p\right)^{\frac{1}{p}}.
\]
Then it is known that
\ben
\item[{\em i)}] $(L^p(\Hcal), ||\cdot||_p)$ is a Banach space,  $L^p(\Hcal)$ is a two-sided ideal
in $\Bcal(\Hcal)$, i.e. for any $a\in L^p(\Hcal)$ and $A,\,B\in \Bcal(\Hcal)$ \,the operator $AaB \in L^p(\Hcal)$
\,and moreover
\[
||AaB||_p \leq ||A||\,||B||\,||a||_p.
\]
\item[{\em ii)}] if $p,q,r\geq 1$ are such numbers that: $\displaystyle \frac{1}{p}+ \frac{1}{q}=\frac{1}{r}$ and 
$a\in L^p(\Hcal)$, $b\in L^q(\Hcal)$, then $a b \in L^r(\Hcal)$.
\item[{\em iii)}] in particular, if $p$ and $q$ satisfy $\displaystyle \frac{1}{p}+ \frac{1}{q}= 1$
then for  $a\in L^p(\Hcal)$, $b\in L^q(\Hcal)$ products  $a b, \ b a \in L^1(\Hcal)$, $\Tr\,ab = \Tr\,ba$  and
\[
|\,\Tr\,a b \,| \leq ||a||_p\,||b||_q.
\]
\een
{\bf Remark.} The space $L^1(\Hcal)$ is the space of trace-class operators on $\Hcal$
and $L^2(\Hcal)$ is the space of Hilbert-Schmidt class. Clearly $L^2(\Hcal)$ equipped
with the sesquilinear form
\[
L^2(\Hcal) \times L^2(\Hcal) \ni (a,b) \longmapsto (a|b) := \Tr\,a^*b \in \Cdb
\]
is a Hilbert space.\\

In what follows we shall also need Hilbert-Schmidt operators in more general settings, namely
the operators from one Hilbert space to another.\\

{\bf Definition.} An operator $c: \Lcal \to \Rcal$ is a Hilbert-Schmidt one, if for
some orthonormal basis $\{\al\}$ in $\Lcal$, the sum
\[
||c||^2_2 := \sum_\alpha (c \,\al | c\,\al) = \sum_\al ||c\al||^2
\]
is finite.

The set of such operators will be denoted by $L^2(\Lcal, \Rcal).$
Clearly for $c\in L^2(\Lcal, \Rcal)$ and any orthonormal basis $\{\beta\}$ in $\Rcal$, we have
\[
\sum_\alpha (c \, \al | c\,\al)
=
\sum_\alpha ( \al | c^* c\, \al)
=
\sum_{\al,\bet}|(\bet | c\, \al)|^2
\]
\be
=
\sum_{\al,\bet} |(c^*\beta|\al)|^2
=
\sum_\bet ||c^*\bet||^2
=
\sum_\beta ( \beta | c\,c^*\beta).
\ee
Therefore  $c^* \in  L^2(\Rcal, \Lcal)$ and the finiteness condition does not depend on the
particular choice of an orthonormal basis $\{\al\}$. In particular  $c^*c\in L^1(\Lcal), \ cc^*\in L^1(\Rcal)$ i.e. they
are trace-class operators acting on $\Lcal$ and $\Rcal$ respectively. Moreover
\be
||c||^2_2 = ||c^*\,c||_1 = \Tr_\Lcal (c^*\,c) = \Tr_\Rcal (c\,c^*) = ||c\,c^*||_1 = ||c^*||^2_2.
\label{c2}
\ee
Let us note that for $a,\,b \in L^2(\Lcal, \Rcal)$ we have $a^*\,b \in L^1(\Lcal), \ b\,a^* \in L^1(\Rcal)$.
Moreover one can easily check that for any Hilbert spaces  $\Lcal', \Lcal, \Rcal, \Rcal'$ a modified ideal property holds:
\be
\left( \begin{array}{c}
 \ c\in L^2(\Lcal, \Rcal) \ \\
    {\rm and} \\
    \ A\in \Bcal(\Lcal',\Lcal), \ B\in \Bcal(\Rcal, \Rcal') \\
    \end{array} \right) \ \ \Rightarrow  
\left(\, B\,c\,A\in L^2(\Lcal', \Rcal')\,\right).
\label{cideal}
\ee
As before the space $L^2(\Lcal, \Rcal)$ forms a Hilbert space equipped with the scalar product
\be
(a| b):= \Tr_\Lcal(a^* \,b) = \Tr_\Rcal(b\,a^*) = (b^*|a^*).
\ee
The last equality can be verified by the similar calculation as above. {\em  In what follows to simplify notation the corresponding
indices $\Lcal$ or $\Rcal$ will be omitted.}
As the result, by  the Schwarz inequality followed by mean arithmetic-geometric inequality,  we obtain
{\begin{col} For arbitrary $a,b \in L^2(\Lcal, \Rcal)$ we have
\be
|\Tr\,a^*b\,|\leq \frac{1}{2}\,[\,\Tr\,a^*a \,+\, \Tr\,b^*b\,]
\label{SchwPlusAG}
\ee
\end{col}

Now we are ready to prove our result. Assume that $c\in L^2(\Lcal, \Rcal) .$ \ Therefore  $c^*\,c$ and $c\,c^*$ are
 trace-class i.e. $|c| \in L^2(\Lcal)$ and $|c^*| \in L^2(\Rcal)$ and this proves the first part of the theorem.
The second part easily follows from (\ref{cideal}). 

To prove the inequality let us note that by the polar decomposition
theorem $c$ is of the form $\displaystyle  c=u\,|c| $
for the unique partial isometry $u \in \Bcal(\Lcal,\Rcal), \ u: |c|(\Lcal) \rightarrow c(\Lcal)$ \
such that $u^*\,u$ and $u\,u^*$ are projections on the initial and final domain respectively.
Now, by uniqueness of the polar decompostion and  functional calculus of bounded, self-adjoint operators we obtain 
(in the same way as for matricies):
\be
u\sqrt{|c|} = \sqrt{|c^*|}\,u.
\label{ucLcRu}
\ee
Therefore
\be
c = \sqrt{|c^*|}\,u\,\sqrt{|c|}.
\label{cRucL}
\ee

Clearly $\sqrt{|c|} \in \Lcal^4(\Lcal)$ and $\sqrt{|c^*|}\in \Lcal^4(\Rcal)$ and this observation
enables us to follow the proof given for matrices in \cite{KLS2}, \cite{Schupp}.
As in (\ref{PQM})  we define operators $P, Q \in \Bcal(\Lcal)$:
\be
P = u^*\,\sqrt{|c^*|}\,B^*\, \sqrt{|c^*|}\,u,\;\;\;\;\;
Q = \sqrt{|c|}\,A^* \,\sqrt{|c|}.
\label{PQ}
\ee
Remembering that $L^p$-spaces are ideals and  using  property $ii)$ of Schatten ideals (for $p=q=4 $)  we see that $Q \in L^2(\Lcal)$ and  
$\sqrt{|c^*|}\,B^*\,\sqrt{|c^*|}\in L^2(\Rcal)$;  by (\ref{cideal}) we have  $P\in L^2(\Lcal)$ . Now (\ref{SchwPlusAG}) reads 
\be
|\Tr\,P^*Q|\leq \frac{1}{2}[\,\Tr\,P^*P + \Tr\,Q^*Q \,]
\label{SchwPlusPQ}
\ee

To compute the left hand side of the above expression let us notice that $u^*\sqrt{|c^*|} = \sqrt{|c|}\,u^*$
by formula (\ref{ucLcRu}). Using this fact and (\ref{cRucL}) we get
\[
P^*Q = u^*\,\sqrt{|c^*|}\,B\, \sqrt{|c^*|}\,u\,\sqrt{|c|}\,A^* \,\sqrt{|c|} = 
\sqrt{|c|}\,u^*\,B\, c\,A^* \,\sqrt{|c|}.
\]
The operator $u^*\,B\, c\,A^*$ belongs to $L^2(\Lcal)$  due to the modified ideal property (\ref{cideal}).
Since $\sqrt{|c|}\in L^4(\Lcal)$ the operator $\sqrt{|c|}\,u^*\,B\, c\,A^* \in  L^{\frac{4}{3}}(\Lcal)$ by
property $ii)$ of Schatten ideals. Now using property $iii)$ in the case $p=\frac{4}{3}$ and $q=4$ we have
\[
\Tr\,P^*\,Q = \Tr\, \left(\sqrt{|c|}\,u^*\,B\, c\,A^* \,\sqrt{|c|}\right) = \Tr\,(|c|\,u^*\,B\, c\,A^*)   = \Tr \,(c^*\,B\, c\,A^*)
\]
due to (\ref{cstarM}). The module of it coincides with the left hand side of the inequality (\ref{MainIneq}). \\
In the similar manner we compute the right hand side of (\ref{SchwPlusPQ}).
\[
P^*\,P = u^*\,\sqrt{|c^*|}\,B\, \sqrt{|c^*|}\,u\,u^*\,\sqrt{|c^*|}\,B^*\, \sqrt{|c^*|}\,u = u^*\,\sqrt{|c^*|}\,B\, |c^*|\,B^*\, \sqrt{|c^*|}\,u.
\]
Therefore
\[
\begin{array}{rl}
\Tr \,P^*\,P & = \Tr\,\left( u^*\,\sqrt{|c^*|}\,B\, |c^*|\,B^*\, \sqrt{|c^*|}\,u\right) = \left|\left|\sqrt{|c^*|}\,B^*\, \sqrt{|c^*|}\,u\,\right|\right|^2_2\\
                 \mbox{}&\\
                & =  \left|\left|u^*\,\sqrt{|c^*|}\,B\, \sqrt{|c^*|}\,\right|\right|^2_2  = \Tr \,\left(\sqrt{|c^*|}\,B^*\, \sqrt{|c^*|}\,u\,u^*\,\sqrt{|c^*|}\,B\, \sqrt{|c^*|}\,\right)\\
                \mbox{}&\\
                &= \Tr \,\left(\sqrt{|c^*|}\,B^*\, |c^*|\,B\, \sqrt{|c^*|}\,\right)
\end{array}
\]
by (\ref{c2}). Now $\sqrt{|c^*|}\,B^*\in L^4(\Rcal)$ and  $|c^*|\,B\,\sqrt{|c^*|} \in  L^{\frac{4}{3}}(\Lcal)$ and using property $iii)$
again we obtain
\be
\Tr \,P^*\,P =  \Tr \,(|c^*|\,B\, |c^*|\,B^*).
\label{PstarP}
\ee
By the similar reasoning  we get
\be
\Tr \,Q^*\,Q =  \Tr \,\left(\sqrt{|c|}\,A\, |c|\,\,A^*\,\sqrt{|c|}\,\right) = \Tr \,({|c|}\,A\,|c|\,A^*).
\label{QstarQ}
\ee
Combining (\ref{PstarP}) and (\ref{QstarQ}) we have the right hand side of inequality (\ref{MainIneq}). The proof is done.

\dowl

\subsection{Main inequality and expectation  values}
\newcommand{\hilsm}{L}
\newcommand{\lewap}{\mathcal{L}}
\newcommand{\prawap}{\mathcal{R}}
\newcommand{\lewab}{\Gamma}
\newcommand{\lewyind}{\gamma}
\newcommand{\prawab}{\Omega}
\newcommand{\prawyind}{\omega}
\newcommand{\bound}{\mathcal{B}}
\newcommand{\domain}{D}
\newcommand{\mt}{\otimes}
\newcommand{\skal}{\,|\,}
\renewcommand{\Tr}{{\rm Tr}}
In this subsection the main inequality (\ref{MainIneq}) will be expressed in terms of expectation values  of operators 
acting on $\lewap\mt \prawap$, where $\lewap$ and $\prawap$ are separable, infinite dimensional Hilbert spaces.
In that form it will be used in following sections.

\noindent 
Let  $\lewab:=\{\psi_\lewyind\}$ denote a fixed  orthonormal basis in  $\lewap$. It defines a linear map:
$$\tilde{\lewab}: \hilsm^2(\lewap,\prawap)\ni c\mapsto \tilde{\lewab}(c)=\sum\psi_\lewyind\mt c\psi_\lewyind\in \lewap\mt \prawap$$
Basic properties of this map are described by
\begin{lem} The map $\tilde{\lewab}$ is unitary, moreover  
for $B\in \bound(\prawap)$ and $a,c\in \hilsm^2(\lewap,\prawap)$  we have:
$$\tilde{\lewab}(Bc)=(I\mt B)\tilde{\lewab}(c)$$
\begin{equation}\label{ItB}
\left(\tilde{\lewab}(a) \skal (I\mt B)\tilde{\lewab}(c)\right)=\left(\tilde{\lewab}(a) \skal \tilde{\lewab}(Bc)\right)=\Tr_\lewap(a^*Bc)=\Tr_\prawap(Bca^*)=\Tr_\prawap(ca^*B)
\end{equation}
\end{lem}
{\bf Proof:\,} It is straightforward to verify  that $\tilde{\lewab}$ is an isometry i.e.  for  $c_1,c_2\in \hilsm^2(\lewap,\prawap)$ we have 
 $\displaystyle \left(\tilde{\lewab}(c_1)| \tilde{\lewab}(c_2)\right)=\Tr_\lewap(c_1^*c_2)$. 
The formula for $\tilde{\lewab}(Bc)$ is clear. To show unitarity, notice that any  $h\in \lewap\mt \prawap$ is of the form 
$h=\sum\psi_\lewyind\mt r_\lewyind$ for the unique family of vectors $(r_\lewyind)$ in $\prawap$. 
Let us  define a linear map $\tilde{\lewab}_1(h): \lewap \rightarrow \prawap$ by $\tilde{\lewab}_1(h)\psi_\lewyind:=r_\lewyind$. 
Then $\tilde{\lewab}_1(h)\in \hilsm^2(\lewap,\prawap)$ and 
simple calculation shows that $\displaystyle \tilde{\lewab}\tilde{\lewab}_1(h)=h$ and $\tilde{\lewab}_1\tilde{\lewab}(c)=c$, 
 so $\tilde{\lewab}$ is unitary and $\tilde{\lewab}_1=\tilde{\lewab}^*$. The formula (\ref{ItB}) is also clear;
the last two equalities  follow from properties of trace: $\Tr_\lewap(a^*c)=\Tr_\prawap(ca^*)$.
\dowl

\noindent
The  basis $\lewab$ defines also an  antiunitary involution $J_\lewab:\lewap\rightarrow \lewap$:
\begin{equation}\label{JLambda}
J_\lewab \left(\sum k_\lewyind\psi_\lewyind\right):=\sum \overline{k}_\lewyind\psi_\lewyind\,,k_\lewyind\in \Cdb\,\,;\,\,\,\,\,\,\,J_\lewab^*=J_\lewab\,\,;\,\,\,\,\,\,J_\lewab^2=I.
\end{equation}
Note that for $A\in \hilsm^1(\lewap)$ we have:
\begin{equation}
\begin{split}
 \Tr_\lewap(J_\lewab A J_\lewab) & =\sum \left(\psi_\lewyind \skal J_\lewab A J_\lewab \psi_\lewyind\right)= \sum \left(A J_\lewab \psi_\lewyind \skal J_\lewab \psi_\lewyind\right)= 
 \sum \left(A\psi_\lewyind \skal \psi_\lewyind\right)=\\
&= \Tr_\lewap(A^*) 
\end{split}
\end{equation}
%
In the same way for a basis   $\prawab=\{\phi_\prawyind\}$ in  $\prawap$ we have the mapping $\tilde{\prawab}$
$$\tilde{\prawab}: \hilsm^2(\prawap,\lewap)\ni d\mapsto \tilde{\prawab}(d)=\sum d \phi_\prawyind\mt \phi_\prawyind\in \lewap\mt\prawap;$$
with the corresponding antiunitary involution $J_\prawab: \,\,
\displaystyle J_\prawab \left(\sum l_\prawyind\phi_\prawyind\right):=\sum \overline{l_\prawyind}\phi_\prawyind$,  and 
\begin{lem} The map $\tilde{\prawab}$ is unitary, moreover for  $A\in \bound(\lewap)$, $B\in\hilsm^1(\prawap)$ and $b,d,\in \hilsm^2(\prawap,\lewap)$:
$$\tilde{\prawab}(A d)=(A\mt I)\tilde{\prawab}(d)$$
\begin{equation}\label{AtI}
(\tilde{\prawab}(d)\skal(A\mt I)\tilde{\prawab}(b))=  \left(\tilde{\prawab}(d)\skal \tilde{\prawab}(A b)\right)=\Tr_\prawap(d^*A b)=\Tr_\lewap(A b d^*)=\Tr_\lewap(b d^* A)
\end{equation}
$$\Tr_\prawap(J_\prawab B J_\prawab)=\Tr_\prawap(B^*)$$ 
\end{lem}
\dowl

The choice of bases in  $\lewap$ and $\prawap$ gives us all of these objects and the following straightforward lemma describes relations 
between both structures:
\begin{lem}\label{lambda-gamma}
Let $c\in\hilsm^2(\lewap,\prawap)$ and $d\in\hilsm^2(\prawap,\lewap)$.  
Then the following equalities hold: 
\begin{equation}
\begin{array}{rrr}
\tilde{\lewab}(c) =  \tilde{\prawab}(J_\lewab c^* J_\prawab)\,,& \hspace{3ex}&\,\tilde{\lewab}(J_\prawab d^* J_\lewab) = \tilde{\prawab}(d)
\end{array}\end{equation}
\begin{equation}
\begin{array}{rrl}
\left(\tilde{\prawab}(d) \skal \tilde{\lewab}(c)\right) = & \left(\tilde{\prawab}(d) \skal \tilde{\prawab}(J_\lewab c^* J_\prawab)\right)= 
& \Tr_\prawap(d^* J_\lewab c^* J_\prawab)= \\
= & \left(\tilde{\lewab}(J_\prawab d^* J_\lewab) \skal \tilde{\lewab}(c)\right)=&
\Tr_\lewap(J_\lewab d J_\prawab c )\label{ilo-gamma-lambda}
\end{array}\end{equation}
\end{lem}
\dowl

Clearly, the choice of  bases $\lewab,\prawab$ is equivalent  to the choice of a basis $\lewab$ and a 
unitary operator $U:\lewap\rightarrow \prawap$
(the equality  $U(\psi_\lewyind)=\phi_\lewyind$ is a definition of operator $U$ or a basis  $\prawab=\{\phi_\lewyind\}$).
It is sometimes more convenient to use pair $(\tilde{\lewab}, U)$ instead of  $(\tilde{\lewab}, \tilde{\prawab})$.
In the lemma below we collect formulae we will use:
\begin{lem}\label{formula-bounded} 
For   $c\in\hilsm^2(\lewap,\prawap)$, $A\in\bound(\lewap)$, $B\in\bound(\prawap)$ 
the following equalities hold:
\begin{align}
\label{jot}
J_\prawab & =U  J_\lewab U^*\\
\label{ItB-mod}
\left(\tilde{\lewab}(c) \skal (I\mt B)\tilde{\lewab}(c)\right)& =\left(\tilde{\lewab}(|c^*|U) \skal (I\mt B)\tilde{\lewab}(|c^*|U)\right) \\
\label{AtI-mod}
\left(\tilde{\lewab}(c) \skal (A\mt I)\tilde{\lewab}(c)\right) & =
\left(\tilde{\lewab}( U |c| ) \skal (A\mt I) \tilde{\lewab}(U |c| )\right)\\
\label{lewa-prawa}
\left(\tilde{\lewab}(c) \skal (A\mt I)\tilde{\lewab}(c)\right) & =
\left(\tilde{\lewab}(U|c|) \skal (I\mt  U J_\lewab A^* J_\lewab U^*)\tilde{\lewab}(U|c|)\right)\\
\label{prawa-lewa}
\left(\tilde{\lewab}(c) \skal (I\mt B)\tilde{\lewab}(c)\right) & =
\left(\tilde{\lewab}(|c^*| U) \skal (J_\lewab U^* B^* U J_\lewab\mt I)\tilde{\lewab}(|c^*|U )\right)\\
\label{lewa-strona-nier}
\left(\tilde{\lewab}(c) \skal (A\mt B)\tilde{\lewab}(c)\right) &=\Tr_\lewap(c^* B c J_\lewab A^* J_\lewab )
\end{align}

\end{lem}
{\bf Proof: } For   $c\in \hilsm^2(\lewap,\prawap)$ recall that   
$|c|:=\sqrt{c^* c}\in \hilsm^2(\lewap)$
and 
$|c^*|=\sqrt{c c^*}\in \hilsm^2(\prawap)$;
notice also that $U|c|\in \hilsm^2(\lewap,\prawap)$ and $|c^*| U\in \hilsm^2(\lewap,\prawap)$.

\noindent The formula (\ref{jot}) is straightforward,  let us prove (\ref{ItB-mod}).  For $B\in \bound(\prawap)$, compute:
\begin{equation*}
\begin{split}
\left(\tilde{\lewab}(c) \skal (I\mt B)\tilde{\lewab}(c)\right) & = \Tr_\lewap(c^*Bc)= \Tr_\prawap(c c^*B)=\Tr_\prawap(|c^*|^2B)= \Tr_\prawap(|c^*| B |c^*|) =\\
& =\Tr_\lewap(U^* |c^*| B |c^*| U)=\Tr_\lewap((|c^*|U)^* B |c^*|U)=\\
&=\left(\tilde{\lewab}(|c^*|U) \skal (I\mt B)\tilde{\lewab}(|c^*|U)\right)
\end{split}
\end{equation*}
For the next formula, let  $A\in\bound(\lewap)$ and using  lemma \ref{lambda-gamma} we compute: 
\begin{equation*}
\begin{split}
\left(\tilde{\lewab}(c) \skal (A\mt I)\tilde{\lewab}(c)\right) & =
\left(\tilde{\prawab}(J_\lewab c^* J_\prawab) \skal (A\mt I)\tilde{\prawab}(J_\lewab c^* J_\prawab)\right)=
\Tr_\prawap((J_\lewab c^* J_\prawab)^* A J_\lewab c^* J_\prawab)=\\
& =\Tr_\prawap(J_\prawab c J_\lewab A (J_\lewab c^* J_\prawab))=
\Tr_\lewap(J_\lewab c^* J_\prawab J_\prawab c J_\lewab A )=\\
&= \Tr_\lewap(J_\lewab |c|^2 J_\lewab A )
\end{split}
\end{equation*}
Writing the formula above for  $U|c|$ instead of $c$ and noting  that  $|U|c||^2=(U|c|)^* U|c|=|c|^2$ we obtain the equality (\ref{AtI-mod}).

\noindent We prove the equality (\ref{lewa-prawa}):
\begin{equation*}\begin{split}
\left(\tilde{\lewab}(c) \skal (A\mt I)\tilde{\lewab}(c)\right) & =
\Tr_\lewap(J_\lewab |c|^2 J_\lewab A )=\Tr_\lewap((J_\lewab |c|J_\lewab) J_\lewab |c| J_\lewab A )=\Tr_\lewap( J_\lewab |c| J_\lewab A J_\lewab |c|J_\lewab)=\\
& =\Tr_\lewap(|c| J_\lewab A^* J_\lewab |c|)=\Tr_\lewap((U|c|)^* U J_\lewab A^* J_\lewab U^* (U|c|))=\\
& =\left(\tilde{\lewab}(U|c|) \skal (I\mt  U J_\lewab A^* J_\lewab U^*)\tilde{\lewab}(U|c|)\right)
\end{split}
\end{equation*}
\noindent  and the formula (\ref{prawa-lewa}):
\begin{equation*}
\begin{split}
\left(\tilde{\lewab}(c) \skal (I\mt B)\tilde{\lewab}(c)\right)& = \Tr_\prawap(|c^*| B |c^*|)=\Tr_\prawap(J_\prawab |c^*| B^* |c^*| J_\prawab)=\\
& = \Tr_\prawap(J_\prawab |c^*| U J_\lewab J_\lewab U^* B^* U J_\lewab J_\lewab U^*  |c^*| J_\prawab)=\\
&=\Tr_\prawap((J_\lewab U^* |c^*| J_\prawab)^*(J_\lewab U^* B^* U J_\lewab) (J_\lewab U^*  |c^*| J_\prawab))=\\
& =\left(\tilde{\prawab}(J_\lewab U^* |c^*| J_\prawab) \skal (J_\lewab U^* B^* U J_\lewab\mt I)\tilde{\prawab}(J_\lewab U^*  |c^*| J_\prawab)\right)=\\
& = \left(\tilde{\lewab}(|c^*| U) \skal (J_\lewab U^* B^* U J_\lewab\mt I)\tilde{\lewab}(|c^*|U )\right)
\end{split}
\end{equation*}
Finally, we prove (\ref{lewa-strona-nier}):
\begin{equation*}
\begin{split}
\left(\tilde{\lewab}(c) \skal (A\mt B)\tilde{\lewab}(c)\right)& =\left((A^*\mt I)\tilde{\lewab}(c) \skal (I\mt B)\tilde{\lewab}(c)\right)=
\left((A^*\mt I)\tilde{\prawab}(J_\lewab c^* J_\prawab) \skal \tilde{\lewab}(B c)\right)=\\
& =\left(\tilde{\prawab}(A^* J_\lewab c^* J_\prawab) \skal \tilde{\lewab}(B c)\right)=\Tr_\prawap(J_\prawab c J_\lewab  A J_\lewab  c^* B^*J_\prawab)=\\
& = \Tr_\prawap(B c J_\lewab  A^* J_\lewab  c^*)=\Tr_\lewap(J_\lewab A^* J_\lewab c^* J_\prawab J_\prawab B c)=\\
&=\Tr_\lewap(J_\lewab A^* J_\lewab c^* B c)=\Tr_\lewap(c^* B c J_\lewab A^* J_\lewab ),
\end{split}
\end{equation*}
where we have used (\ref{ilo-gamma-lambda}).

\dowl

Now we want to express  the inequality (\ref{MainIneq}) i.e.
\begin{equation}\label{nierownosc}
2 |\Tr_\lewap(c^* B c A^*)|\leq \Tr_\lewap(|c|A |c|A^*)+\Tr_\prawap(|c^*|B|c^*| B^*)
\end{equation}
in terms of $\tilde{\lewab},J_\lewab, U$ and the scalar product in $\lewap\mt \prawap$. 

\noindent
In the formula (\ref{lewa-strona-nier}) we put $B:=U J_\lewab AJ_\lewab U^*$ and $U|c|$ instead of $c$ and get:
\begin{equation*}
\begin{split}
\left(\tilde{\lewab}(U|c|) \skal (A\mt (U J_\lewab A J_\lewab U^*))\tilde{\lewab}(U|c|)\right) & =
\Tr_\lewap(|c| U^*(U J_\lewab A J_\lewab U^*)(U |c|) J_\lewab A^*J_\lewab)=\\
 & =\Tr_\lewap(|c| J_\lewab A J_\lewab |c| J_\lewab A^*J_\lewab)
\end{split}
\end{equation*}
Now put into (\ref{lewa-strona-nier}) $A=U^*J_\prawab B J_\prawab U$ and  $|c^*|U$  instead  of $c$ and obtain:
\begin{equation*}
\begin{split}
\left(\tilde{\lewab}(|c^*| U) \skal (U^*J_\prawab B J_\prawab U \mt B)\tilde{\lewab}(|c^*|U)\right)& =
\Tr_\lewap(U^* |c^*| B |c^*| U  J_\lewab (U^*J_\prawab B J_\prawab U)^* J_\lewab )=\\
=\Tr_\lewap(U^* |c^*| B |c^*| (U  J_\lewab U^*) J_\prawab B^* J_\prawab U  J_\lewab ) & =
\Tr_\lewap(U^* |c^*| B |c^*| (J_\prawab J_\prawab) B^* J_\prawab (J_\prawab U))=\\
& =\Tr_\lewap(U^* |c^*| B |c^*|  B^*  U)=\\
&= \Tr_\prawap(|c^*| B |c^*|  B^*),
\end{split}
\end{equation*}
where (\ref{jot}) i.e. $J_\prawab=U J_\lewab U^*$ was used.

\noindent
Finally, writing (\ref{nierownosc}) with $J_\lewab A J_\lewab$ instead of  $A$, using (\ref{lewa-strona-nier}) and two equalities above
we obtain:
\begin{prop}
Let $A\in \bound(\lewap)$, $B\in \bound(\prawap)$ and $c\in \hilsm^2 (\lewap,\prawap)$. Then 
\begin{equation}\label{nier-inter}
\begin{split}
2 \left| \left(\tilde{\lewab}(c) \skal (A\mt B)\tilde{\lewab}(c)\right)\right|\,\, & \leq  \,\,\left(\tilde{\lewab}(U|c|) \skal (A\mt U J_\lewab A J_\lewab U^*)\tilde{\lewab}(U|c|)\right)+\\
& \hspace{6ex} + \left(\tilde{\lewab}(|c^*| U) \skal (U^*J_\prawab B J_\prawab U \mt B)\tilde{\lewab}(|c^*|U)\right)
\end{split}
\end{equation}
\end{prop}

\noindent
We will also  need formulae  similar to the ones in Lemma \ref{formula-bounded} for some unbounded operators, they are proven in the following proposition.
\begin {prop}\label{unb}
Let $T$ and $S$ be  self-adjoint operators with purely point spectrum acting on $\lewap$ and $\prawap$ respectively. Assume that bases
$\{\psi_\lewyind\}$ and $\{\phi_\lewyind\} $ consist of eigenvectors of $T$ and $S$:  $T\psi_\lewyind=t_\lewyind\psi_\lewyind$ and  
$S\phi_\lewyind=s_\lewyind\phi_\lewyind$;  assume moreover that 
$\tilde{\lewab}(c)\in \domain((T\mt I)^2)$ and  $\tilde{\prawab}(d)\in \domain((I\mt S)^2)$. Then:
\begin{eqnarray}
\label{unb123}\left(\tilde{\lewab}(c) \skal (T\mt I)\tilde{\lewab}(c)\right)=\left(\tilde{\lewab}(U|c|) \skal (T\mt I)\tilde{\lewab}(U|c|)\right)=
\left(\tilde{\lewab}(U |c|) \skal (I\mt UTU^*)(\tilde{\lewab}(U |c|)\right)\\
\label{unb456}\left(\tilde{\prawab}(d) \skal (I\mt S)\tilde{\prawab}(d)\right)=\left(\tilde{\prawab}(U^*|d|) \skal (I \mt S)\tilde{\prawab}(U^*|d|)\right)=
\left(\tilde{\prawab}(U^* |d|) \skal  (U^* S U \mt I)\tilde{\prawab}(U^* |d|)\right)
\end{eqnarray}
\end{prop}
{\bf Proof: } We will prove  (\ref{unb123}); equalities in (\ref{unb456}) can be proven in a similar manner.\\
Since  $\tilde{\lewab}(c)$ is in the domain of $(T\mt I)^2$ we have:
\begin{equation}
\label{unb1-1}
\begin{split}
(\tilde{\lewab}(c) \skal (T\mt I)^2\tilde{\lewab}(c)) & =\sum_\lewyind (\psi_\lewyind \mt c\psi_\lewyind \skal (T\mt I)^2\tilde{\lewab}(c))=
\sum_\lewyind (t_\lewyind^2 \psi_\lewyind \mt c\psi_\lewyind \skal \tilde{\lewab}(c))=\\
&=  \sum_\lewyind |t_\lewyind|^2 ||c\psi_\lewyind||^2.
\end{split}
\end{equation}
This equality means that the series $\sum_\lewyind t_\lewyind \psi_\lewyind\mt c\psi_\lewyind$ is convergent. Because 
$T\mt I$ is closed and $\psi_\lewyind\mt c\psi_\lewyind\in \domain(T\mt I)$ it implies that:
\begin{equation}
\label{unb1} (T\mt I)\,\tilde{\lewab}(c)=\sum_\lewyind t_\lewyind \psi_\lewyind\mt c\psi_\lewyind
\end{equation}
Since 
$||U|c|\psi_\lewyind||^2=
||c\psi_\lewyind||^2$ the formula (\ref{unb1-1}) implies also convergence of the series $\sum_\lewyind t_\lewyind \psi_\lewyind\mt U|c|\psi_\lewyind$
and the equality 
\begin{equation}
\label{unb2}(T\mt I)\,\tilde{\lewab}(U|c|)=\sum_\lewyind t_\lewyind \psi_\lewyind\mt U|c|\psi_\lewyind
\end{equation}

\noindent By the lemma \ref{lambda-gamma}: $\tilde{\lewab}(U |c|)=\tilde{\prawab}(J_\lewab |c|U^* J_\prawab)$. Since 
$J_\lewab |c| U^*\phi_\lewyind\mt\phi_\lewyind=J_\lewab |c|\psi_\lewyind\mt\phi_\lewyind\in \domain(I\mt U T U^*)$ 
the formula (\ref{unb1-1}) means convergence of the series $\sum_\lewyind t_\lewyind J_\lewab |c| \psi_\lewyind \mt \phi_\lewyind$ 
and the  equality 
\begin{equation}
\label{unb3}(I\mt UTU^*)\,\tilde{\lewab}(U |c|)=\sum_\lewyind t_\lewyind (J_\lewab |c| \psi_\lewyind) \mt \phi_\lewyind
\end{equation}
follows.
Now  combining (\ref{unb1}), (\ref{unb2}), (\ref{unb3}) and  
$\tilde{\lewab}(U |c|)=\tilde{\prawab}(J_\lewab |c|U^* J_\prawab)$  we obtain (\ref{unb123}).

\dowl
\section{Ground state ordering in the system of 2-dimensional rotors}
\label{sec:LRO}
\newcommand{\ekin}{{\rm T}}
\subsection{Description of the system}
\label{subsec:Ham}
Denote by $\La$  the finite subset of the simple cubic lattice in $d$ dimensions:
 $\La\subset \Zdb^d$.
We assume that  $\La$ is a (discrete) hypercube and that {\em the number of sites along every edge is even}; let us fix $2N$ to be
the length of the hypercube edge:
\be
\label{def-sieci}
\Lambda:=\{\xgr\in \Zdb^d: -N+1\leq x_i\leq N \,,\,i=1,\dots,d\}
\ee
With every site $\xgr\in\La$ we associate a real variable $\varphi_\xgr\in[0,2\pi[$. 
In physical terms, it describes the (angular) position of the rotor at the site $\xgr$.
Equivalently, the position of the rotor at the site $\xgr$ can be described as a unit 
vector $\sgr_\xgr \in \mathbb{S}^1$, i.e. one dimensional torus $\Tdb$: 
\[
\sgr_\xgr =
[s_\xgr^x,s_\xgr^y]=[\cos\varphi_\xgr, \sin\varphi_\xgr].
\]
The total spin is  $\,\,\displaystyle \Sgr=\sum_{\xgr\in\La}\sgr_\xgr.$
The Hilbert space $\Hcal_\xgr$ of states on a given site $\xgr$ is the space of square integrable
periodic functions, i.e.   $\displaystyle \Hcal_\xgr = L^2(\Tdb)$. 
The Hilbert space associated to the whole system is the space of square integrable functions on $|\Lambda|$ - dimensional torus:
\[
\Hcal_\La
= \displaystyle L^2(\Tdb^{|\Lambda|}
)=\mathop{\otimes}_{\xgr\in\La}\Hcal_\xgr
\]
The operator $T$ of total kinetic energy of the system of  rotors is proportional to the laplacian $\Delta$:
\be
\ekin=-\frac{1}{2I}\sum_{\xgr\in\La}\frac{\prt^2}{\prt \varphi_\xgr^2}
\label{TotalT},
\ee
where $I>0$ is the moment of inertia of rotor (we assume that all rotors have equal moments of inertia). 
The system of interacting rotors is  defined by the 
Hamiltonian $ H=\ekin+\hat{V}$, 
where $\hat{V}$ is an interaction energy between rotors; it is an operator of multiplication by a smooth function $V$.
We shall consider the Hamiltonian:
\be
H=\ekin+\hat{V_0}\,\,,\,\,\,\,\,
V_0:=- J \sum_{\brak \xgr\ygr\kket}\cos(\varphi_\xgr-\varphi_\ygr)
\label{HamRot}
\ee
In this formula $\brak \xgr\ygr\kket$ means that $\xgr$ and $\ygr$ are {\em the nearest neighbours.
By this we mean that all but one coordinates of $\xgr$ and $\ygr$ are equal and the ones, say $x_i$ and $y_i$, that differ satisfy
$\displaystyle |x_i-y_i|=1\,{\rm mod}\, 2(N-1)$.
Thus a site $\xgr$ laying on the hyperplane  defined by $x_i=N$ has some of its nearest neighbours on a hyperplanes defined by $x_i=-N+1$.}

$J$ is the coupling constant: $J>0$ corresponds to ferromagnetic coupling between rotors and $J<0$ to  the
antiferromagnetic one.  
{\em In what follows we shall restrict ourselves to  the ferromagnetic  case, as only in this situation the Reflection
Positivity arguments may be applied. }

Our Hamiltonian (\ref{HamRot}) is {\em an elliptic second order differential operator} on a $|\Lambda|$-dimensional torus (compact manifold).
It is a special case of the more general situation:
\begin{tw}\label{ellop}\cite{Nicola}
Let $(M,g)$ be a compact, oriented, riemannian smooth  manifold without boundary; 
$L:C^\infty(M)\rightarrow C^\infty(M)$ formally selfadjoint, linear, elliptic, PDO of order $k>0$.
Then:
\begin{enumerate}
\item $L$ extends uniquely to $\tilde{L}: H^k(M)\rightarrow L^2(M)$; ($H^k(M)$ is $k$-th Sobolev space)
\item $\tilde{L}$ as an operator on $L^2(M)$ (with the domain $H^k(M)$) is selfadjoint;
\item The spectrum of $\tilde{L}$ consists of isolated eigenvalues of finite multiplicity;
\item Eigenvectors of $\tilde{L}$ are smooth functions.
\end{enumerate}
\end{tw}

Let us observe that since $\ekin$ is positive and  $V_0$ is a continuous function, the hamiltonian $H$ is {\em bounded from below.}

In the following we need  in an essential way the {\em uniqueness of  the ground state of the Hamiltonian.} 
This is a consequence of  {\em positivity improving property} of  the semigroup $\exp(-t\Delta)$. 
For convenience of the reader, we recall  briefly main definitions and results (we refer to Chapt. XIII of \cite{RS4} for a detailed presentation).

A {\em non zero} function $\Psi\in L^2(M)$  is {\em positive} iff $\Psi(x)\geq 0$; it is {\em strictly positive} if $\Psi(x)>0$ 
(both inequalities should be understood in  almost everywhere sense).\\
A bounded operator $A$ is: \\
\hspace*{2ex}-- {\em positivity preserving} if $A\Psi$ is positive for positive $\Psi$; \\
\hspace*{2ex}--  {\em positivity improving} if $A\Psi$ is strictly positive for positive $\Psi$; equivalent condition is that $(\Psi\skal A\Phi)>0$ for 
positive $\Psi$ and $\Phi$.\vspace{0.5em}

\noindent The following is,  simplified  for our needs,  Thm XIII.44 from \cite{RS4}.
\begin{prop} 
Let $H$ be a self adjoint, bounded from below operator on $L^2(M)$. Assume the  spectrum of $H$ consists   of isolated eigenvalues of finite multiplicity. 
If for every $t>0$ the operator $\exp(-t H)$ is positivity improving then the ground state of $H$ is unique (and strictly positive).
\end{prop}

In our situation $M=\Tdb^{|\Lambda|}$, and it is known that for kinetic energy operator $\ekin$ given by the formula (\ref{TotalT}) 
operators $\exp(-t\ekin)\,,\,t>0$ are integral operators;  for $\Phi,\,\Psi\in L^2(\Tdb^{|\Lambda|})$ :
$$\left(\Phi\skal \exp(-t\ekin) \Psi\right)=\int dx d y \overline{\Phi(x)} K(t,x,y) \Psi(y).$$
The function $K(t,x,y)\,,\,t>0, x,y\in \Tdb^{|\Lambda|}$ -- the heat kernel for $|\Lambda|$ - torus, is explicitely known; it is {\em a strictly positive} function, 
therefore $\exp(-t\ekin)$ is positivity improving.

The same property for $H=\ekin+\hat{V}$ holds due to the Trotter product formula (see e.g Chapt. X of \cite{RS2}):
$$\exp(-t(\ekin +\hat{V}))\Psi=\lim_{n\rightarrow\infty}\left[\exp(-t \ekin /n) \exp(-t\hat{V}/n)\right]^n\Psi$$
%
In fact, for any $\Phi,\,\Psi\in \Hcal_\La$ : 
$$(\Phi \skal \exp(-t H) \Psi)=\lim_{n\rightarrow\infty}\left(\Phi\skal  \left[\exp(-t \ekin /n) \exp(-t\hat{V}/n)\right]^n \Psi\right)$$
Let $C:=\sup V$. Since $\exp(-t\ekin)$ is positivity improving, for strictly positive numbers: $\alpha,\beta,\gamma,\delta>0$ and positive functions 
$\Phi$ and $\Psi$  we have:
\begin{eqnarray}
\left(\Phi\skal  \exp(-\alpha \ekin) \exp(-\beta \hat{V}) \exp(-\gamma \ekin) \exp(-\delta \hat{V})\Psi\right)\geq 
e^{-\beta C} \left(\exp(-\alpha \ekin) \Phi\skal \exp(-\gamma \ekin) \exp(-\delta \hat{V})\Psi\right)\geq \nonumber\\
\geq e^{-(\beta +\delta) C}\left( \Phi\skal \exp(-(\alpha+\gamma) \ekin) \Psi\right).\mbox{ \hspace{1.8cm} } \nonumber
\end{eqnarray}
In particular    
$\ \ \displaystyle \left(\Phi\skal  \left[\exp(-t \ekin /n) \exp(-t\hat{V}/n)\right]^n \Psi\right)\geq  e^{-t C} \left(\Phi\skal \exp(-t \ekin ) \Psi\right)> 0$.

\noindent
Therefore  the inequality is preserved in the limit: \ $(\Phi \skal \exp(-t H) \Psi)>0$ and the uniqueness of the ground state of $H$ follows.

\subsection{Criteria of ordering}

The simplest definition of the order parameter would be an average of the total spin. However, this definition is of little use for the zero field (i.e. as
a measure of the spontaneous magnetization) as it is zero due to symmetry. The more physical definition
is a zero-field limit of magnetization: $\Mcal =\limf{h}{0} M(h) $ ($h$ denotes magnetic field)
but it is difficult to deal with. More easy to handle is the {\em average of the square of spin}. 
It follows that if the average
of the square of spin is different from zero, then the zero-field magnetization is non-zero, too
(Griffiths theorem -- see \cite{DLS}). So, we take the average $\brak \Sgr^2\kket$ as a measure of
order parameter. {\em  All averages considered in this paper are taken over the ground state}.
Following this idea, we will prove that if $I$ and $J$ are sufficiently large 
the ground state of system of interacting rotors, described by the potential  (\ref{HamRot}),
exhibits Long-Range Order (LRO):
\begin{tw}
\label{tw2}
Assume that $I$ and $J$ satisfy the inequality
\be
\sqrt{I J} > \frac{1}{(2\pi)^d}
\int_{[-\pi,\pi]^d} \frac{\df\kgr }{\sqrt{\Ecal(\kgr)}}
\equiv \Ical_d,
\label{MainIneqDoTh2},
\ee
where the function $\Ecal: \Rdb^d \rightarrow \Rdb$ is defined by:
\be
\Ecal(\kgr) = d-\sum_{i=1}^d\cos k_i \label{Epsilon}.
\ee
Then there exists $C>0 $ such that, for sufficiently large $|\La|$: 
\be
 \left\langle\left(\frac{\Sgr^x}{|\La|}\right)^2 \right\rangle
\equiv
\left\langle\left(\frac{1}{|\La|}\sum_{\xgr\in \La} s_\xgr^x\right)^2\right\rangle
\equiv
\left\langle\left(\frac{1}{|\La|}\sum_{\xgr\in \La} \cos\varphi_\xgr \right)^2\right\rangle
\geq C,
\label{Ordering}
\ee
\end{tw}

{\em Remarks.} 
\ben
\item The estimation of the form similar to the one in (\ref{Ordering}) appeared 
in P--based proofs of LRO in other classes of models  including 
classical and quantum spin systems \cite{DLS}, \cite{FSS76}, \cite{PaChor}.
\item The integral $\Ical_1$ is divergent, so above theorem does not 
prove existence of ground-state LRO in $d=1$ but  scaling and field-theoretical 
arguments \cite{Sachdev}  show  (non-rigorously) that there is no ordering in this case.
\item For $d\geq 2$ the integral $\Ical_d$ is finite , so  there is  LRO in the ground-state.  Numerical values of $\Ical_d$ for physical dimensions  
are: $\Ical_2\approx 0.909173; \Ical_3\approx 0.643954$. It indicates the  tendency to  ordering  increases with the growth of a dimension.
\item
For  positive temperatures, there is no ordering In dimensions 1 and 2.
This statement, analogous to the famous
Mermin-Wagner theorem for spin systems \cite{MerminWagner}, has been 
proven in \cite{PaChor}.
\item It is known that in $d\geq 3$,
the LRO is present in sufficiently low temperatures, and  so in the 
ground state, too \cite{PaChor}.  Therefore the Theorem \ref{tw2} is most 
interesting in $d=2$. Such a  result, using another RP arguments, has been proven in \cite{W1}. In different approach, by 
scaling and field-theoretical arguments, it was obtained (non-rigorously)  in \cite{Sachdev}.
\een

\subsection{Estimations for basic functions}
\label{subsec:WazneFunkcje}
To prove the theorem \ref{tw2} it will be  convenient to work with Fourier-transformed spins. Let us define:
\be
\hat{s}_\kgr^\al = \frac{1}{\sqrt{|\La|}} \sum_{\xgr\in\La}s_\xgr^\al e^{i\kgr\cdot\xgr}
\;\;\;\;\;
(\al=x,y)
\label{SpinFourier}
\ee
where $\kgr$ takes value in the first Brillouin zone, i.e. 
$k_j\in\{-\frac{\pi(N-1)}{N},\dots,\frac{\pi(N-1)}{N},\pi\}$ for $j=1,\dots,d$.
%
Let us remark that due to the symmetry of our system we have:
\begin{lem} For the system of rotors described by the hamiltonian (\ref{HamRot}) operators $\hat{s}_\kgr^\al$ ($\alpha=x,y$) satisfy identities:
\be \label{spin-prop}
\left\langle \hat{s}^\al_\kgr \right\rangle=0\,\,\,\,\,{\rm and} \,\,\,\,
\left\langle
\hat{s}^x_\kgr (\hat{s}^x_{\kgr})^*
\right\rangle =
\left\langle
\hat{s}^y_\kgr (\hat{s}^y_{\kgr})^*
\right\rangle.\ee
\end{lem}
{\em Proof: }  Let $R_\theta$ be the rotation (in all variables) by $\theta$: 
$$(R_\theta f)(\varphi_1,\dots,\varphi_{|\Lambda|}):=f(\varphi_1-\theta ,\dots,\varphi_{|\Lambda|}-\theta)$$
It is clear  that $R_\theta$ is unitary.  Since  the   hamiltonian (\ref{HamRot}) commutes with $R_\theta$ and {\em the ground state is unique}
we have $R_\theta\psi_0=\lambda \,\psi_0$ for a complex number $\lambda$ with $|\lambda|=1$. Therefore
$$\left\langle A \right\rangle=\left(\psi_0\skal A \psi_0\right)= \left(R_\theta \psi_0\skal A R_\theta \psi_0\right)=\left\langle R_\theta^* A R_\theta\right\rangle$$
It is easy to check that for any  $\kgr$:  $ R_\pi\hat{s}^\al_\kgr=-\hat{s}^\al_\kgr R_\pi$ and $R_{\pi/2}\hat{s}^x_{\kgr}=\hat{s}^y_{\kgr}R_{\pi/2}$.
Now equalities (\ref{spin-prop}) are  clear.
\dowl

Let us denote by $g_\kgr$ the two-point correlation function in the momentum representation:
\be
g_\kgr =
\left\langle
\hat{s}^x_\kgr (\hat{s}^x_{\kgr})^*
\right\rangle
\label{DefOf_gk}
\ee
Clearly $g_\kgr\geq 0$. Since  $\hat{s}^x_{\kgr}$ is normal $g_\kgr=\left\langle(\hat{s}^x_\kgr)^* \hat{s}^x_{\kgr}\right\rangle$.   Remember also that  $g_\kgr$ 
depends on parameters $I$ and $J$.
{\em The $g_\kgr$ function and its estimation will play a crucial role in the proof of 
existence of spontaneous magnetization. }

With this notation the inequality (\ref{Ordering}) can be rewritten as 
\begin{equation}
\frac{1}{|\Lambda|}g_\0gr\geq C>0.
\end{equation}
The strategy of the proof of Theorem \ref{tw2} can be described as follows (general ideas are similar to ones in  \cite{FSS76}, \cite{DLS}, \cite{KLS2}).
First (and rather easy) step is the equality
\be
\label{suma_g}
\sum_{\kgr} g_\kgr=\frac{1}{2}\,|\La|
\ee
Indeed, for the groud state $\psi_0$ of $H$:
$$\sum_\kgr g_\kgr=\sum_\kgr(\hat{s}^x_\kgr \psi_0\skal \hat{s}^x_{\kgr} \psi_0)= 
\sum_{\kgr, \xgr, \ygr} \frac{1}{|\Lambda|}(\cos \varphi_\xgr e^{i \kgr\xgr}\psi_0\skal  \cos \varphi_\ygr e^{i \kgr\ygr}\psi_0 )=
\sum_\xgr (\cos^2 \varphi_\xgr \psi_0\skal \psi_0 ), $$
where  we have used the obvious formula:
$\displaystyle \sum_\kgr e^{i  \kgr(\xgr-\ygr)}=|\Lambda| \delta(\xgr-\ygr)$.
In the same way we obtain:
$\sum_\kgr(\hat{s}^y_\kgr \psi_0\skal \hat{s}^y_{\kgr} \psi_0)=
\sum_\xgr (\sin^2 \varphi_\xgr \psi_0\skal \psi_0 ).$
Adding these two  equalities we get:
$$\sum_{\kgr}
\left\langle
\hat{s}^x_\kgr (\hat{s}^x_{\kgr})^*
\right\rangle +
\left\langle
\hat{s}^y_\kgr (\hat{s}^y_{\kgr})^*
\right\rangle=|\Lambda|
$$
and the equality (\ref{suma_g}) follows due to the second formula of (\ref{spin-prop})


\noindent
It turns out, that for $\kgr\ne \0gr$, {\em and this is the place where the RP arguments and the main inequality (\ref{nier-inter}) is used},  the function $g_\kgr$ 
can be estimated from above by an integrable (for $d>1$) function: 
\be
g_\kgr\leq
\frac{1}{2 \sqrt{IJ} \sqrt{\Ecal(\kgr)}} ,
\;\;\;\;\;\kgr\ne \0gr.
\label{g_estimation}
\ee
where the function $\Ecal(\kgr)$ was defined in (\ref{Epsilon}). 

\noindent
With (\ref{suma_g}) and (\ref {g_estimation}) in hand we can write
$$
\frac{1}{|\La|} g_\0gr =\frac{1}{2}-\sum_{\kgr\neq 0} g_\kgr\geq
\frac{1}{2\sqrt{I J}}
\left(\sqrt{I J}-\frac{1}{|\La|} \sum_{\kgr\ne \0gr}\frac{1}{\sqrt{\Ecal(\kgr)}}\right)$$
Now the sum  $\displaystyle \frac{(2 \pi)^d}{|\Lambda|}\sum_{\kgr\ne\0gr} \frac{1}{\sqrt{\Ecal(\kgr)}}$ converges  as $\Lambda\rightarrow \Zdb^d$ to the integral
$\displaystyle \int_{[-\pi,\pi]^d} \frac{\df\kgr}{\sqrt{\Ecal(\kgr)}}.$
Therefore the inequality (\ref{MainIneqDoTh2}) 
implies that there exists $C>0$ such that for sufficiently large $\Lambda$ we have
$$\left(\sqrt{I J}-\frac{1}{|\La|} \sum_{\kgr\ne \0gr}\frac{1}{\sqrt{\Ecal(\kgr)}}\right)\geq C$$
i.e the estimate (\ref{Ordering}). {\em This way to complete the proof of the theorem \ref{tw2} it remains  to prove the estimate (\ref {g_estimation})}.
The existence of such an estimate seems to be a quite general phenomenon, but at present we can prove it only using  RP techniques.



The inequality  (\ref{Ordering}) can be viewed as the appearance of the macroscopic occupation of the $\kgr = \0gr$ mode. In the other words, it is an 
indication that in the thermodynamic limit, the $g_\kgr$ function possess non-zero $\del$ function contribution at $\kgr=0$. 

We will get the  inequality (\ref{g_estimation}) by relating $g_\kgr$ to other functions, in particular {\em susceptibility} (\ref{chiByR}), for which we will get an 
estimate by RP techniques. 

Let $(\psi_n)_{n\geq 0}$ be an orthonormal basis consisting of eigenvectors of the Hamiltonian (\ref{HamRot}) with $H\psi_n=E_n\psi_n$ and $\psi_0$ be the ground state.
We have:
\begin{equation}
\label{gk1}
\begin{split}
g_\kgr & = \left\langle(\hat{s}^x_{\kgr})^*\hat{s}^x_\kgr \right\rangle=\left( \hat{s}^x_\kgr \psi_0\skal \hat{s}^x_\kgr \psi_0\right)=
\sum_{n\geq 0} \left( \hat{s}^x_\kgr \psi_0\skal \psi_n\right) \left( \psi_n\skal \hat{s}^x_\kgr \psi_0\right)=\\
& = \sum_{n>0} \left( \hat{s}^x_\kgr \psi_0\skal \psi_n\right) \left( \psi_n\skal \hat{s}^x_\kgr \psi_0\right)=
\sum_{n>0} \left|\left( \hat{s}^x_\kgr \psi_0\skal \psi_n\right)\right|^2.
\end{split}
\end{equation}
There is no term with $n=0$ due to the first formula  of (\ref{spin-prop}).
Since  $\hat{s}^x_\kgr$ is normal $g_\kgr$ can be writtes as:
\be \label{gk2}
g_\kgr=\left\langle\hat{s}^x_\kgr  (\hat{s}^x_{\kgr})^*\right\rangle=\sum_{n>0} \left|\left( (\hat{s}^x_\kgr)^* \psi_0\skal \psi_n\right)\right|^2=
\frac12\sum_{n>0} \left[\left|\left( \hat{s}^x_\kgr \psi_0\skal \psi_n\right)\right|^2+ 
\left|\left( (\hat{s}^x_\kgr)^* \psi_0\skal \psi_n\right)\right|^2\right]\ee

\noindent
Next, we define:
\be\label{Dk-def}
\Dcal_\kgr :=
\frac{1}{2}\left\langle
[[\hat{s}^x_\kgr, H], (\hat{s}^x_{\kgr})^*]
\right\rangle\ee
Since  $(\hat{s}^x_\kgr) \psi_0$ and $(\hat{s}^x_\kgr)^* \psi_0$ are in the domain of $H$ this definition is correct
Short calculation shows that:
\be
\Dcal_\kgr=\frac12\sum_{n>0} \left[\left|\left( \hat{s}^x_\kgr \psi_0\skal \psi_n\right)\right|^2+ 
\left|\left( (\hat{s}^x_\kgr)^* \psi_0\skal \psi_n\right)\right|^2\right] (E_n-E_0)
\label{DCbyR}
\ee
\noindent

\noindent
We shall also use the {\em susceptibility} which is given by:
\be
\chi_\kgr =\frac12\sum_{n>0} \left[\left|\left( \hat{s}^x_\kgr \psi_0\skal \psi_n\right)\right|^2+ 
\left|\left( (\hat{s}^x_\kgr)^* \psi_0\skal \psi_n\right)\right|^2\right] \frac{1}{E_n-E_0}
\label{chiByR}
\ee
Clearly $\chi_\kgr<\infty$ due to (\ref{gk2}) and the fact that $\lim E_n= + \infty$.

\noindent
For a positive integer $n>0 $ let 
$$a_n:=\sqrt{\left|\left( \hat{s}^x_\kgr \psi_0\skal \psi_n\right)\right|^2+ 
\left|\left( (\hat{s}^x_\kgr)^* \psi_0\skal \psi_n\right)\right|^2} \frac{1}{\sqrt{E_n-E_0}}\,,
$$ 
$$b_n:=\sqrt{\left[\left|\left( \hat{s}^x_\kgr \psi_0\skal \psi_n\right)\right|^2+ 
\left|\left( (\hat{s}^x_\kgr)^* \psi_0\skal \psi_n\right)\right|^2\right] (E_n-E_0)}\,.
$$
Then $g_\kgr^2=\left(\sum a_n b_n\right)^2$ and (by Schwarz inequality):
$$g_\kgr^2=\left(\sum a_n b_n\right)^2\leq \left(\sum a_n^2\right)\,\left(\sum b_n^2\right) =\chi_\kgr \Dcal_\kgr$$

Notice, that since $\hat{s}^x_\kgr$ is a multiplication by a smooth function and $H$ is a second order differential operator, the double
commutator appearing in (\ref{Dk-def}) is an operator of multiplication by a smooth function, so it is bounded. In fact by a direct calculation one gets
\be
\Dcal_\kgr=\frac{1}{2}\left\langle
[[\hat{s}^x_\kgr, H], (\hat{s}^x_{\kgr})^*]
\right\rangle
\leq
\frac{1}{4I}.
\label{EstimOfDCdlaT0}
\ee
Using  this estimate 
we obtain the inequality:
\be\label{gkchik} g^2_\kgr\leq \chi_\kgr \cdot \Dcal_\kgr\leq \frac{\chi_\kgr}{4 I}.\ee

\noindent
{\em This shows that an upper  bound for $\chi_\kgr$ implies the  upper  bound for $g_\kgr$.}\vspace{1ex}
\subsection{Reflection Positivity arguments} 
Let us now proceed along the general line of RP arguments.
We perturb the Hamiltonian in the analogous manner as it was done
in the case of positive temperatures
\cite{PaChor}, \cite{W1}.
To do it, 
let us  first modify the original Hamiltonian (\ref{HamRot})  by a constant (so irrelevant) term introduced by  the potential:
\be
V= \frac{J}{2} 
\sum_{\brak \xgr\ygr\kket}
\left[
(\cos\varphi_\xgr-\cos\varphi_\ygr)^2 + (\sin\varphi_\xgr-\sin\varphi_\ygr)^2
\right]=V_0 + J  
\sum_{\brak \xgr\ygr\kket} 1
\label{HamRot2}
\ee
Now, for a function  $b: \Lambda \ni \xgr\mapsto b_\xgr\in \Cdb$ defined on sites, let us consider the perturbed potential $V(b)$:
\be
V(b)= 
 \frac{J}{2}
\sum_{\brak \xgr\ygr\kket}
 \left[
\left|\cos\varphi_\xgr - b_\xgr-\cos\varphi_\ygr+ b_\ygr\right|^2 + (\sin\varphi_\xgr-\sin\varphi_\ygr)^2
\right]
\ee
and the perturbed hamiltonian
\be\label{HamRotGnrlzd}
H(b):=\ekin+\hat{V}(b). 
\ee
Let us remark, that if functions $b$ and $b'$ differ by a constant function then $H(b)=H(b')$, in particular for $b$ being a constant $H(b)=H(0)$.

Clearly the perturbed Hamiltonian satisfies all assumptions of Thm  \ref{ellop}. 
We will use the Reflection Positivity and the operator inequality in the form given in (\ref{nier-inter}) to prove the following:
\begin{tw}Let $E_0(b)$ be the ground state energy of the perturbed hamiltonian (\ref{HamRotGnrlzd}). Then:
\be
E_0(b)\geq E_0(0)=:E_0
\label{EhGeqE0}
\ee
\end{tw}

The rest of the  subsection is devoted to the proof of above theorem.   
Let us recall the definition of $\Lambda$ (\ref{def-sieci}):
\be
\Lambda:=\{\xgr\in \Zdb^d: -N+1\leq x_i\leq N \,,\,i=1,\dots,d\}
\ee
We divide the system into two identical subsystems  $\La_L,\La_R$ so that $\La=\La_L\cup\La_R$; 
$\La_L$ is an mirror image of $\La_R $ under reflection in the $\Pi$ (hyper) plane given by the equation $x_1=\frac12.$
This way,  the subset $\La_L$ contains all sites,
where the first coordinate is negative or $0$, and  $\La_R$ -- sites, where it  is positive.

\begin{figure}
        \centering
                \includegraphics{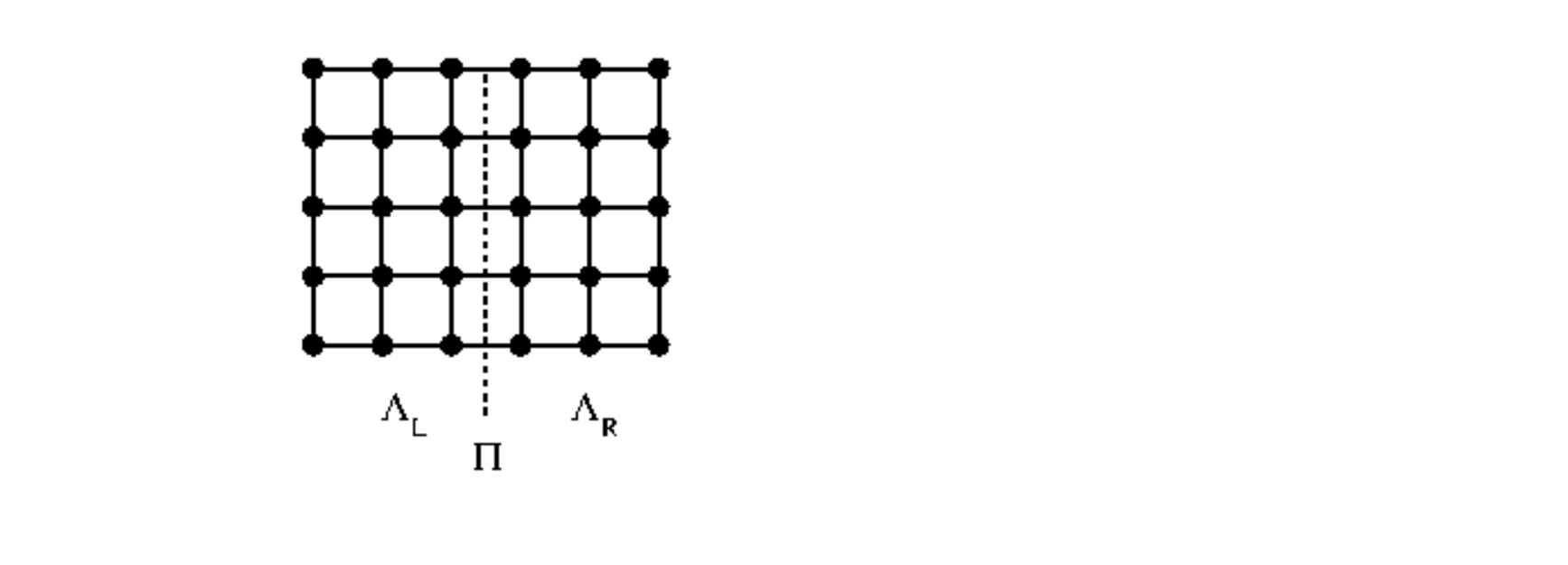}
        \caption{ \scriptsize Division of the system into two identical subsystems 
$\La_L, \La_R$ by the symmetry plane $\Pi$. 
The illustration concerns open boundary conditions.
Let us stress that in the paper we have to do with periodic boundary conditions, where analogous division can be made.
}
        \label{fig:RP}
\end{figure}

The  Hilbert space $\Hcal_\La$ of states of the whole  system is a tensor product of two spaces:
 $\Hcal_\La = \lewap  \otimes \prawap $, where $\lewap$ ($\prawap$)
  is a space of states of subsystem defined on 
$\La_L$ ($\La_R$).

\noindent
For a function $b:\Lambda\rightarrow \Cdb$ define:
\be
\left\{
\begin{array}{ccccccc}
b_L = b & {\rm on} & \La_L, &   b_L & \mbox{is defined on} & \La_R & \mbox{by 
the mirror symmetry}\\
b_R = b & {\rm on} & \La_R, &   b_R & \mbox{is defined on} & \La_L & \mbox{by 
the mirror symmetry}\\
\end{array}
\right.
\label{DefsOf_hL_and_hR}
\ee
(the standard trick in RP). Our first step to prove inequality (\ref{EhGeqE0}) is the following :
\begin{lem}\label{nier-bl-br}
Let $b:\Lambda \rightarrow \Cdb$ be a function and $b_L, b_R$ be related to $b$ as in  (\ref{DefsOf_hL_and_hR}).
Let  hamiltonians $H(b),H(b_R), H(b_L)$ be defined  by (\ref{HamRotGnrlzd}) and  $E_0(b), E_0(b_R),  E_0(b_L)$ denote their ground state  energies respectively. Then:
\be2 E_0(b)\geq E_0(b_R) + E_0(b_L)\label{EhEhREhL}\ee
\end{lem}
We move the proof of this lemma to the end of this subsection.\vspace{1ex}

Now, we will apply RP arguments to show how the inequality  (\ref{EhEhREhL}) implies (\ref{EhGeqE0}). 
For a given function $b:\Lambda\rightarrow \Cdb$ let us call a {\em non-zero bond for} $b$  a pair of nearest neighbours $(\xgr,\ygr)$ with $b(\xgr)\neq b(\ygr)$.
Clearly, if $b\neq const$ then there are $l>0$ non-zero bonds for $b$ and we can choose the  symmetry plane, which crosses at least 
one of them.  Notice that if  $l_L$, $l_R$ are  number of non-zero bonds for $b_L$ and $b_R$ respectively 
then, we have:
\[
l_L+l_R<2l;
\]
i.e.  at least one of the numbers $(l_L, l_R)$  is  less  than $l$. Therefore for any non-constant $b$ we can pass to a constant one by applying finitely many
replacements  $b\rightarrow b_L$ or $b\rightarrow b_R$ (for different symmetry planes). For a given lattice $\Lambda$ the maximal number of steps is bounded independently of $b$ -- 
let $K$ denotes this bound.

%

\newcommand{\emin}{\mathbf E}
\noindent
Our modified hamiltonian (\ref{HamRotGnrlzd}) is a {\em positive} operator, so the set $ \{E_0(b)\}$ is bounded from below and let us define $\emin:=\inf\{E_0(b)\}$.
From the inequality (\ref{EhEhREhL}) it follows that for $\epsilon>0$:
\be 
{\rm if\,\,}  \emin \leq E_0(b) \leq \emin + \epsilon\,\,\,{\rm then\,\,}\, \emin \leq E_0(b_L) \leq  \emin+2 \epsilon \,
\,{\rm and}\,\,  \emin \leq E_0(b_R) \leq  \emin+2 \epsilon \label{step}
\ee
So let $\epsilon>0$ be given. There exists $\tilde{b}$ with $\displaystyle \emin \leq E_0(\tilde{b}) \leq \emin + 2^{-K} \epsilon$;  replacing succesively $\tilde{b}$ by
$\tilde{b}_L$ or $\tilde{b}_R$ and applying (\ref{step}) we obtain $\tilde{b}_0=const$ with
$$\emin \leq E_0(\tilde{b}_0) \leq \emin + \epsilon;$$
but since $E_0(\tilde{b}_0)=E_0(0)=E_0$ it follows that $\emin=E_0$ -- so to prove the inequality (\ref{EhGeqE0}) it remains to prove the lemma \ref{nier-bl-br}.
\\
{\em Proof of the lemma \ref{nier-bl-br}:}
Consider the perturbed hamiltonian  $\displaystyle H(b)=\ekin+\hat{V}(b)$, where $V(b)$ is given by (\ref{HamRotGnrlzd}).
The proof consists of two steps: the first is rather involved --  by using the lemma (\ref{formula-bounded}) and prop (\ref{unb}) we will show that for a 
{\em smooth} function $\Psi\in\lewap\mt\prawap$, there exist $\Psi_L,\Psi_R\in \lewap\mt\prawap$ such that:
\be 2 \left(\Psi\skal H(b)\Psi\right)\geq \left(\Psi_L \skal H(b_L) \Psi_L\right) + \left(\Psi_R \skal H(b_R) \Psi_R\right)\label{podlemat}\ee
Since, by {\em the variational principle} $\left(\Psi_L \skal H(b_L) \Psi_L\right)\geq E_0(b_L)$ and $\left(\Psi_R \skal H(b_R) \Psi_R\right)\geq E_0(b_R)$,
we get:
$$\left(\Psi_L \skal H(b_L) \Psi_L\right)\geq E_0(b_L) + E_0(b_R).$$
For $\Psi=\psi_0$ -- the ground state of $H(b)$, we obtain the inequality (\ref{EhEhREhL}).  

To simplify notation let us define $a_\xgr(b):=\cos \varphi_\xgr-b_\xgr$; we will write just $a_\xgr$ if it is clear what is $b$. Now, $V(b)$ reads:
$$V(b)=\frac{J}{2}\sum_{<\xgr\ygr>}|a_\xgr-a_\ygr|^2+(\sin\varphi_\xgr-\sin\varphi_\ygr)^2$$
Let us define subsets $B_L\subset \La_L$, $B_R\subset \La_R$:
\be
\begin{array}{ccl}
B_L & := & \{\xgr\in\La_L: \mbox{the first coordinate of }\xgr\mbox{ is } 0 \mbox{ or } -N+1\}\\
B_R & := & \{\xgr\in\La_R: \mbox{the first coordinate of }\xgr\mbox{ is } 1\mbox{ or } N \}\\
\end{array}
\label{BLandBR}
\ee
Notice that $B_L$ ($B_R$) is the subset of those elements in $\La_L$ ($\La_R$) which have (some of) their nearest neighbours in $\La_R$ ($\La_L$). 
The potential $V(b)$ can be written as:
\begin{equation*}
\begin{split}
V(b) & =\frac{J}{2}\sum_{<\xgr\ygr>\subset L}|a_\xgr-a_\ygr|^2+(\sin\varphi_\xgr-\sin\varphi_\ygr)^2 + 
\frac{J}{2}\sum_{<\xgr\ygr>\subset R}|a_\xgr-a_\ygr|^2+(\sin\varphi_\xgr-\sin\varphi_\ygr)^2+ \\
& + \frac{J}{2}\sum_{\xgr\in B_L}|a_\xgr-a_{\xgr'}|^2+(\sin\varphi_\xgr-\sin\varphi_{\xgr'})^2,
\end{split}\end{equation*}
where  $\xgr'$ is the image of $\xgr$ by the reflection across the hyperplane $\Pi$. The last term of the sum above we write as:
%
%
\begin{equation*}
\begin{split}
\frac{J}{2}\sum_{\xgr\in B_L}|a_\xgr-a_{\xgr'}|^2+(\sin\varphi_\xgr-\sin\varphi_{\xgr'})^2 & =
 \frac{J}{2}\sum_{\xgr\in B_L}\left(|a_\xgr|^2+ \sin^2\varphi_\xgr\right)
+\frac{J}{2}\sum_{\xgr\in B_L}\left(|a_{\xgr'}|^2+\sin^2\varphi_{\xgr'}^2\right)+\\
& -\frac{J}{2} \sum_{\xgr\in B_L}\left(\overline{a_\xgr}a_{\xgr'}+a_\xgr \overline{a_{\xgr'}}+2 \sin\varphi_\xgr \sin\varphi_{\xgr'}\right)
\end{split}\end{equation*}
Let us define functions:
\begin{equation}\label{VLB}
V_L(b)
:=\frac{J}{2}\sum_{<\xgr\ygr>\subset L}|a_\xgr(b)-a_\ygr(b)|^2+(\sin\varphi_\xgr-\sin\varphi_\ygr)^2+
\frac{J}{2}\sum_{\xgr\in B_L}|a_\xgr(b)|^2+ \sin^2\varphi_\xgr
\end{equation}
\begin{equation}\label{VRB}
V_R(b)
:=\frac{J}{2}\sum_{<\xgr\ygr>\subset R}|a_\xgr(b)-a_\ygr(b)|^2+(\sin\varphi_\xgr-\sin\varphi_\ygr)^2+
\frac{J}{2}\sum_{\xgr\in B_L}|a_{\xgr'}(b)|^2+ \sin^2\varphi_{\xgr'}
\end{equation}
\begin{equation}\label{VIB}
V_{I}(b):=-\frac{J}{2} \sum_{\xgr\in B_L}\left(\overline{a_\xgr}a_{\xgr'}+a_\xgr \overline{a_{\xgr'}}+2 \sin\varphi_\xgr \sin\varphi_{\xgr'}\right)
\end{equation}
Using  above notation we can write the potential $V(b)$ as:
$V(b)=V_L(b)+V_R(b)+V_I(b)$ and the corresponding operator $\hat{V}(b)$ as 
$\hat{V}(b)=\widehat{V_L}(b)\mt I+I\mt \widehat{V_R}(b) +\widehat{V_I}(b)$, where
\begin{equation}\label{VIB-op}
\widehat{V_I}(b):=-\frac{J}{2} \sum_{\xgr\in B_L}\left(\overline{a_\xgr}\mt a_{\xgr'}+a_\xgr \mt \overline{a_{\xgr'}}+2 \sin\varphi_\xgr \mt \sin\varphi_{\xgr'}\right).
\end{equation}
 For the  kinetic term $\ekin$ we have $\ekin=\ekin_L\mt I+ I\mt \ekin_R$.

Now, Let us choose an orthonormal basis $\lewab:=\{\psi_\lewyind\}$ in  $\lewap$ {\em consisting of eigenvectors of $\ekin_L$}: 
$\ekin_L \psi_\lewyind=t_\lewyind \psi_\lewyind$ and  the corresponding involution $J_\lewab$. 
%
%
Note that since $\ekin_L$ is self-adjoint it commutes with $J_\lewab$:
\begin{equation}\label{T-J}
J_\lewab \ekin_LJ_\lewab=\ekin_L.
\end{equation}
Let $U:\lewap\rightarrow \prawap$ be a unitary operator and  $\phi_\lewyind:=U\psi_\lewyind$ be 
the corresponding basis  in $\prawap$. 

Let us also {\em assume} that 
\begin{equation} \label{TLTR} \ekin_R=U \ekin_L U^*\end{equation}
That means, in particular, that $\phi_\lewyind\in\domain(\ekin_R)$ and $\ekin_R\phi_\lewyind=t_\gamma\phi_\lewyind$

Let $\Psi=:\tilde{\Gamma}(c)\in \lewap\mt \prawap$ be a {\em smooth function}; then it belongs to domains of $\ekin^k, (\ekin_L\mt I)^k, (I\mt \ekin_R)^k$ 
for $k=1,2,3,\dots$ and, using (\ref{VLB}), (\ref{VRB}) and (\ref{VIB}),  we can write:
\begin{align}
2 (\Psi\skal H(b)\Psi)
& =2 (\Psi \skal \ekin \Psi)+2 (\Psi\skal \widehat{V}(b))\Psi)=\nonumber\\
& =(\Psi \skal (\ekin_L\mt I) \Psi)+ (\Psi\skal (\ekin_L\mt I) \Psi)+   \label{line2}\\
\label{line3}& +(\Psi\skal (I\mt \ekin_R)\Psi)+(\Psi\skal (I\mt \ekin_R)\Psi)+\\
\label{line4}& + (\Psi\skal (\widehat{V_L}(b)\mt I)\Psi)+ (\Psi\skal (\widehat{V_L}(b)\mt I)\Psi)+ \\
\label{line5}& +(\Psi \skal (I \mt \widehat{V_R}(b))\Psi)+ (\Psi\skal (I \mt \widehat{V_R}(b))\Psi)+\\
\label{line6}& +2 (\Psi\skal \widehat{V_I}(b)\Psi)
\end{align}

\noindent 
Now, due to the  lemma \ref{formula-bounded} and proposition \ref{unb},  we are going to rewrite various terms appearing in this equality in different form.

\noindent By formulae (\ref{unb123}) and (\ref{TLTR}) for terms in (\ref{line2})we have:
\begin{align*}
\left(\Psi \skal (\ekin_L\mt I) \Psi\right) & =\left(\tilde{\lewab}(U |c|) \skal (\ekin_L\mt I)\tilde{\lewab}(U |c|)\right)\\
\left(\Psi\skal (\ekin_L\mt I) \Psi\right) & = 
\left(\tilde{\lewab}(U |c|) \skal (I\mt \ekin_R)\tilde{\lewab}(U |c|)\right);\nonumber
\end{align*}
Using (\ref{unb456}) with $d=J_\lewab c^* J_\prawab$ (then $|d|=J_\prawab|c^*|J_\prawab$ and $\tilde{\prawab}(U^*|d|)=\tilde{\lewab}( |c^*| U)$ ) 
and  (\ref{TLTR}) for terms in (\ref{line3}):
\begin{align*}
(\Psi \skal (I \mt \ekin_R) \Psi) & =\left(\tilde{\lewab}( |c^*|U ) \skal (I\mt \ekin_R)\tilde{\lewab}( |c^*|U)\right)\nonumber\\
(\Psi \skal (I \mt \ekin_R) \Psi) & =\left(\tilde{\lewab}( |c^*|U ) \skal (\ekin_L \mt I)\tilde{\lewab}( |c^*|U)\right)\nonumber
\end{align*}
For terms appearing in (\ref{line4}) and (\ref{line5}) by the use of  (\ref{AtI-mod}), (\ref{lewa-prawa}), (\ref{ItB-mod}) and (\ref{prawa-lewa}) we get:
\begin{align*}
\left(\Psi \skal (\widehat{V_L}(b)\mt I) \Psi\right) & =
\left(\tilde{\lewab}(U |c|) \skal (\widehat{V_L}(b)\mt I)\tilde{\lewab}(U |c|)\right)\\
\left(\Psi \skal (\widehat{V_L}(b)\mt I) \Psi\right)& =
\left(\tilde{\lewab}(U |c|) \skal (I\mt U J_\lewab (\widehat{V_L}(b))^* J_\lewab U^*)\tilde{\lewab}(U|c|)\right)\\
(\Psi \skal (I\mt\widehat{V_R}(b))\Psi)& =\left(\tilde{\lewab}( |c^*|U ) \skal (I\mt\widehat{V_R}(b))\tilde{\lewab}( |c^*|U)\right)\\
(\Psi \skal (I\mt\widehat{V_R}(b))\Psi)& =
(\tilde{\lewab}( |c^*|U ) \skal (U^*J_\prawab(\widehat{V_R}(b))^*J_\prawab U \mt I)\tilde{\lewab}( |c^*|U))
\end{align*}
Now we are going to use the inequality (\ref{nier-inter}) to estimate the term in (\ref{line6}).\\ 
Since $\left(\Psi \skal (\overline{a_\xgr}\mt a_{\xgr'}+a_\xgr \mt \overline{a_{\xgr'}})\Psi\right)$ is  {\em real} we have:
\begin{equation*}
\begin{split}
\left(\Psi \skal (\overline{a_\xgr}\mt a_{\xgr'}+a_\xgr \mt \overline{a_{\xgr'}})\Psi\right)& \leq 
\left|\left(\Psi \skal (\overline{a_\xgr}\mt a_{\xgr'}+a_\xgr \mt \overline{a_{\xgr'}})\Psi\right)\right|\leq \\
& \leq \left|\left(\Psi \skal (\overline{a_\xgr}\mt a_{\xgr'})\Psi\right)\right|+ \left|\left(\Psi \skal (a_\xgr \mt \overline{a_{\xgr'}})\Psi\right)\right|
\end{split}
\end{equation*}
Applying the inequality (\ref{nier-inter}) to  each of two terms we get:
\begin{equation*}
\begin{split}
2 \left|\left(\Psi|(\overline{a_\xgr}\mt a_{\xgr'})\Psi\right)\right|& \leq  
\left(\tilde{\lewab}(U |c|) \skal (\overline{a_\xgr}\mt U J_\lewab \overline{a_\xgr} J_\lewab  U^*)\tilde{\lewab}(U |c|)\right)+\\
& + \left(\tilde{\lewab}(|c^*|U) \skal (U^* J_\prawab a_{\xgr'} J_\prawab U \mt a_{\xgr'})\tilde{\lewab}(|c^*|U)\right)
\end{split}
\end{equation*}
and
\begin{equation*}
\begin{split}
2 \left|\left(\Psi|(a_\xgr \mt \overline{a_{\xgr'}})\Psi\right)\right|& \leq  
\left(\tilde{\lewab}(U |c|) \skal (a_\xgr\mt U J_\lewab a_\xgr J_\lewab  U^*)\tilde{\lewab}(U |c|)\right)+\\
& + \left(\tilde{\lewab}(|c^*|U) \skal (U^* J_\prawab \overline{a_{\xgr'}} J_\prawab U \mt \overline{a_{\xgr'}})\tilde{\lewab}(|c^*|U)\right)
\end{split}
\end{equation*}
Similarly, since $2 \left(\Psi \skal (\sin\varphi_\xgr \mt \sin\varphi_{\xgr'})\Psi\right)$ is  {\em real}:
\begin{equation*}
\begin{split}
2 \left(\Psi \skal (\sin\varphi_\xgr \mt \sin\varphi_{\xgr'})\Psi\right)& \leq 2 \left| \left(\Psi \skal (\sin\varphi_\xgr \mt \sin\varphi_{\xgr'})\Psi\right)\right|\leq\\
& \leq \left (\tilde{\lewab}(U |c|) \skal (\sin\varphi_\xgr \mt U J_\lewab \sin\varphi_\xgr J_\lewab  U^*)\tilde{\lewab}(U |c|)\right)+\\
&+ \left(\tilde{\lewab}(|c^*|U) \skal (U^* J_\prawab \sin\varphi_{\xgr'}J_\prawab  U\mt \sin\varphi_{\xgr'})\tilde{\lewab}(|c^*|U)\right)
\end{split}
\end{equation*}
Adding these three inequalities and multiplting by {\em a negative number} $(-J/2)$ we obtain the estimate for (\ref{line6}):
$$2 (\Psi \skal \widehat{V_I}(b)\Psi)\geq   \left(\tilde{\lewab}(U |c|)\skal  K_L \,\tilde{\lewab}(U |c|)\right)+
\left(\tilde{\lewab}(|c^*|U)\skal K_R\, \tilde{\lewab}(|c^*|U)\right),$$
where operators $K_L$ and $K_R$ are defined  by:
\begin{eqnarray}
\label{kl}
K_L := - \frac{J}{2} \sum_{\xgr\in B_L} 
\left[\overline{a_\xgr}\mt U J_\lewab \overline{a_\xgr} J_\lewab  U^*+ a_\xgr\mt U J_\lewab a_\xgr J_\lewab  U^*+
2 \sin\varphi_\xgr \mt U J_\lewab \sin\varphi_\xgr J_\lewab  U^*\right]\\
\label{kr}
 K_R := - \frac{J}{2} \sum_{\xgr\in B_L} \left[ U^*J_\prawab a_{\xgr'} J_\prawab U \mt a_{\xgr'} +
U^*J_\prawab \overline{a_{\xgr'}} J_\prawab U \mt \overline{a_{\xgr'}} + 2 U^*J_\prawab\sin\varphi_{\xgr'} J_\prawab U \mt \sin\varphi_{\xgr'}\right]
\end{eqnarray}
Putting all together  we obtain the following inequality :
\begin{eqnarray}
2 (\Psi \skal H(b)\Psi)\geq 
\left(\Psi_L \skal H_1 \,\Psi_L \right)+
\left(\Psi_R \skal H_2 \,\Psi_R\right), {\rm\,\,where\,}\\
\Psi_L;=\tilde{\lewab}(U |c|)\,,\,\Psi_R:=\tilde{\lewab}( |c^*|U)\\
\label{h1} H_1:=\ekin_L\mt I + I\mt \ekin _R +\widehat{V_L}(b)\mt I 
+I\mt U J_\lewab (\widehat{V_L}(b))^* J_\lewab U^*+ K_L\\
\label{h2} H_2:=\ekin_L \mt I+ I\mt \ekin _R+ U^*J_\prawab(\widehat{V_R}(b))^*J_\prawab U \mt I + 
I\mt\widehat{V_R}(b)+ K_R
\end{eqnarray}
\noindent
We would like to have   $H_1=H(b_L)\,,\,\, H_2=H(b_R)$.
Notice that the $H_1$ and $H_2$ depend only on composition $U J_\lewab$ (by (\ref{jot}) $J_\prawab U = U J_\lewab $);  
remember also  that we have assumed in (\ref{TLTR}) that $\ekin_R=U \ekin_L U^*.$

Let $M:\Lambda\rightarrow\Lambda$ be the reflection across our (hyper)plane $\Pi$ and let $U_M:\lewap\rightarrow\prawap$ be the unitary defined by
\begin{eqnarray}
(U_M \psi)(\varphi_{\ygr_1},\varphi_{\ygr_2},\dots):=\psi(\varphi_{\xgr_1},\varphi_{\xgr_2},\dots)\\
\ygr_i:=M(\xgr_i)\,,\,\xgr_i\in \Lambda_L\,,\,i=1,\dots,|\Lambda_L|,\,\nonumber
\end{eqnarray}
and $J_0$ be the complex conjugation (we will use $J_0$  for conjugations on $\lewap$ and $\prawap$). It is clear that 
$$J_0 U_M=U_M J_0\,\,, \,\,\ekin_L J_0=J_0 \ekin_L\,\,,\,\, \ekin_R J_0=J_0 \ekin_R,$$
Define  the unitary $U:=J_0 U_M J_\lewab=U_M  J_0 J_\lewab$; then
\be U \ekin_L U^*=U_M  J_0 J_\lewab \ekin_LJ_\lewab J_0 U_M^*= U_M  \ekin_L U_M^*=\ekin_R\, \,{\rm and}\,\,UJ_\lewab=J_0 U_M=U_M J_0.
\ee
We will analyze  $H_1$. 
Clearly we have $V_L(b)=V_L(b_L)$;  since the function $V_L(b)$ is real $(\widehat{V_L}(b))^*=\widehat{V_L}(b)$ and $J_0 \widehat{V_L}(b)J_0=\widehat{V_L}(b)$. Therefore 
$$U J_\lewab (\widehat{V_L}(b))^* J_\lewab U^*=U_M \widehat{V_L}(b) U_M^*=\widehat{V_R}(b_L)$$
This way we  obtain: $$H_1=\ekin+\widehat{V_L}(b_L)\mt I + I\mt \widehat{V_R}(b_L)+K_L$$
Let us  show that $K_L$ given by (\ref{kl}) is equal $\widehat{V_I}(b_L)$ defined in (\ref{VIB-op}):
\begin{equation*}
\begin{split}
U J_\lewab \overline{a_\xgr(b)} J_\lewab  U^*& =U_M J_0 \overline{(\cos\varphi_\xgr-b_\xgr)}J_0 U_M^*=U_M (\cos\varphi_\xgr-b_\xgr) U_M=\cos\varphi_{\xgr'}-b_L(\xgr')=\\
&=a_{\xgr'}(b_L),
\end{split}
\end{equation*}
where $\xgr':=M(\xgr)$ (because for $\xgr\in\Lambda_L$  $b(\xgr)=b_L(\xgr)=b_L(\xgr')$);
the next term:
$$U J_\lewab a_\xgr(b) J_\lewab  U^*=U_M\overline{a_\xgr(b)}U_M^*=\overline{a_{\xgr'}(b_L)}$$
and finally
$$U J_\lewab \sin\varphi_\xgr J_\lewab  U^*=U_M \sin\varphi_\xgr U_M^*=\sin\varphi_{\xgr'}$$
So we get:
$$K_L = - \frac{J}{2} \sum_{\xgr\in B_L} 
\left[\overline{a_\xgr(b_L)}\mt a_{\xgr'}(b_L) + a_\xgr(b_L)\mt \overline{a_{\xgr'}(b_L)} +
2 \sin\varphi_\xgr \mt \sin\varphi_{\xgr'}\right]
$$
and this is $\widehat{V_I}(b_L)$, so really $H_1=H(b_L)$; 
in the same way one gets $H_2=H(b_R).$ 

This way the proof of the   inequality (\ref{podlemat}) and the  lemma \ref{nier-bl-br} is completed as well as the proof of the inequality  (\ref{EhGeqE0}).
\dowl

\subsection{Estimations giving LRO}

In this subsection we finally complete the proof of LRO,  i.e. we show the  inequality  (\ref{g_estimation}).
The inequality (\ref{EhGeqE0})  implies that
\be
\left.\frac{\df^2 E(\la b)}{\df\la^2}\right|_{\la=0}\geq 0
\label{SecDervEb}
\ee
for an {\em arbitrary} $b$. It turns out that if we take $b$ being 
{\em plane wave} with the wave
vector $\kgr$  then we get (\ref{g_estimation}).
Let us present calculations in more details. 

For the moment, let us keep the $b$ function
being arbitrary. Write the perturbed Hamiltonian (\ref{HamRotGnrlzd}) with the $b$ function
rescaled by a factor $\la\in\R$:

\be
H(\la b)
=
H(0) +\la H'(b) +\la^2 C(b),
\label{Hdo2rz}
\ee
where
\be
H'(b):= -J Re\left(\sum_{<\xgr\ygr>} (\cos\varphi_\xgr-\cos\varphi_\ygr)(b_\xgr-b_\ygr)\right)\,,\,\,\,C(b):=\frac{J}{2}\sum_{<\xgr\ygr>}|b_\xgr-b_\ygr|^2
\label{HaPrim}
\ee

\noindent
Let   $E_0(\la b)$ be the ground state energy of the operator $H(\la b)$, and
$\stackrel{(2)}{\Del}E_0(b)$ -- the correction to ground state energy in the second order perturbation
theory for the Hamiltonian $H(0)+\la H'(b)$, i.e.
$$\stackrel{(2)}{\Del}E_0(b)=\sum_{n>0}\frac{|(\psi_n\skal H'(b) \psi_0)|^2}{E_0-E_n}$$
Therefore:
\be
\left.\frac{\df^2 E(\la b)}{\df\la^2}\right|_{\la=0}=
\stackrel{(2)}{\Del}E_0(b)+2 C(b) \geq 0.
\label{SecDervEbInaczej}
\ee
Choose now the $b$ function  as
\be
\bsf_\xgr = \frac{1}{\sqrt{|\La|}} e^{i\kgr\cdot \xgr}
\label{bPW}
\ee
With such a choice (\ref{HaPrim}) reads: 
\be
C(b):=\frac{J}{2|\La|}\sum_{<\xgr\ygr>}|e^{i\kgr\cdot \xgr}-e^{i\kgr\cdot \ygr}|^2=J\Ecal(\kgr)
\ee
and, since
\be
\frac{1}{\sqrt{|\La|}}\sum_{\brak \xgr\ygr\kket} (\cos\varphi_\xgr-\cos\varphi_\ygr)
( e^{i\kgr\cdot \xgr}- e^{i\kgr\cdot \ygr})
= s^x_\kgr\sum_{j=1}^d(2-2\cos \kgr_j) \equiv
2  s^x_\kgr \Ecal(\kgr),
\ee 
$$H'(b)=- 2 J \Ecal(\kgr) Re(s^x_\kgr)\,,\,\,$$
Therefore by (\ref{SecDervEbInaczej}):
$$-4J^2 \Ecal^2 (\kgr) \sum_{n>0}\frac{|(\psi_n\skal Re(s^x_\kgr) \psi_0)|^2}{E_n-E_0}+2 J\Ecal(\kgr)\geq 0$$
By the similar computations, replacing $b$ by $i b$ we obtain:
$\displaystyle H'(i b)= 2 J \Ecal(\kgr) Im (s^x_\kgr)$ and 
$$-4J^2 \Ecal^2 (\kgr) \sum_{n>0}\frac{|(\psi_n\skal Im(s^x_\kgr) \psi_0)|^2}{E_n-E_0}+2 J\Ecal(\kgr)\geq 0$$
Adding these inequalities we get (remember $J \Ecal(\kgr) \geq 0$):
$$\sum_{n>0}\frac{|(\psi_n\skal Re(s^x_\kgr) \psi_0)|^2}{E_n-E_0}+\frac{|(\psi_n\skal Im(s^x_\kgr) \psi_0)|^2}{E_n-E_0}\leq \frac{1}{ J \Ecal(\kgr)}.$$
For  complex numbers $\alpha:=(\psi_n\skal s^x_\kgr \psi_0)$ and $\beta:=(\psi_n\skal (s^x_\kgr)^* \psi_0)$, by the paralleogram law :
$|\alpha+\beta|^2+|\alpha-\beta|^2=2(|\alpha|^2+|\beta|^2$, we get
$$|(\psi_n\skal Re(s^x_\kgr) \psi_0)|^2+|(\psi_n\skal Im(s^x_\kgr) \psi_0)|^2=\frac{1}{2}\left( (\psi_n\skal s^x_\kgr \psi_0)|^2+|(\psi_n\skal (s^x_\kgr)^* \psi_0)|^2\right),$$
and finally: 
$$\frac{1}{2}\sum_{n>0}\frac{|(\psi_n\skal s^x_\kgr \psi_0)|^2}{E_n-E_0}+\frac{|(\psi_n\skal (s^x_\kgr)^* \psi_0)|^2}{E_n-E_0}\leq \frac{1}{J \Ecal(\kgr)}.$$
The LHS of this inequality is just succeptibility $\chi_\kgr$, compare (\ref{chiByR}),  so   we have 
$\chi_\kgr\leq  \frac{1} {J \Ecal(\kgr)}$.
Using (\ref{EstimOfDCdlaT0}) and (\ref{gkchik}), we obtain the estimate (\ref{g_estimation}):
\be
g^2_\kgr\leq \chi_\kgr \cdot \Dcal_\kgr\leq \chi_\kgr \frac{1}{4 I}\leq \frac{1}{4 I J \Ecal(\kgr)}\nonumber
\ee
The proof of theorem \ref{tw2} is complete.

\dowl

\section{Summary}
\label{sec:Summary}

We have extended the Kennedy-Lieb-Shastry-Schupp matrix inequality to 
the case where matrices
are replaced by certain  infinite dimensional operators.  Similar result 
has been proven in \cite{PaChor} for
another matrix inequality -- the DLS lemma, which is crucial for the 
proof of occurrence of LRO in the system of
interacting rotors in low temperatures in $d\geq 3$.

With the use of this inequality and Reflection Positivity technology, we 
have formulated
sufficient condition (\ref{MainIneqDoTh2}) for ordering in the ground 
state of the system of
interacting rotors. In particular,  the LRO is present in
$d\geq 2$ for sufficiently large value of $IJ$. This way, we have shown 
the occurrence of the LRO
in the ground state of interacting rotor systems in a direct manner. In 
the paper \cite{W1}, analogous
result has been proven by Reflection Positivity technique,
  but without checking some assumptions (validity of certain limiting 
procedure). Our present
approach does not suffer from this drawback. This result has also been 
obtained in non-rigorous way by scaling and field-theoretic arguments 
\cite{Sachdev}

We are convinced that our result can be extended to other rotor systems: 
other (bipartite) lattices and
larger space of internal degrees of freedom, for instance, for $O(n)$ 
systems.

One can pose the problem concerning the occurrence
  the ordering in  opposite situation, i.e.
for quantity $IJ$ being small. To our best knowledge, this is an open 
question. One can
suspect that the LRO should be absent. Such expectation is motivated by 
the paper
  \cite{VerbZagr}, where somewhat similar result has been proved:
There is no ordering in  the anharmonic
crystal model provided mass of the oscilator is sufficiently small.

There are numerous interesting rotor-like systems, which do {\em not} 
fulfill conditions
allowing an application of Reflection Positivity techniques. One of most 
important of them,
is the lattice system of interacting bosons (for instance, the Bose 
Hubbard model). The
Hamiltonian of this system, written in the language of coherent states, 
becomes the Hamiltonian
of interacting rotors of the form (\ref{HamRot}), plus one term more 
(see, for instance \cite{BoseHumCS}). This last term spoils
the Reflection Positivity, an it seems to be not possible to apply these 
techniques to the
analysis of interacting boson systems. (Only exception is the paper 
\cite{ALSSY}, where the
Bose-Einstein condensation has been proved  for hard-core bosons on 
optical lattice. Here, the
term spoiling RP is absent due to the hard-core condition). Here we
tackle with the long-standing and important problem: How to extend the 
range of applicability
of Reflection Positivity technique, which works for certain problems, 
and does not work for
apparently very similar ones.

Another interesting problem is the occurrence of the Kosterlitz-Thouless 
transition \cite{KT} in the two-dimensional
rotor system. On physical grounds, one can expect occurrence of this 
transition, at least for large momentum of
inertia (the quantum-mechanical rotors should not differ too much from 
the 2d XY model, for which such
a transition has been rigorously proven \cite{FrohlichSpencer}) However, 
we are not aware on rigorous results for interacting
rotor systems.

\vskip.3cm

{\bf Acknowledgments.} We thank Przemys\l aw Majewski for discussions on early stages of this work.


\end{document}